\documentclass[12pt]{article}
\usepackage{graphicx} 
\usepackage{amsmath}
\usepackage[letterpaper,top=1in,bottom=1in,left=1in,right=1in]{geometry}
\usepackage[utf8]{inputenc}
\usepackage[T1]{fontenc}
\usepackage{textcomp}
\usepackage{newtxtext}
\usepackage{newtxmath}
\usepackage{microtype}
\usepackage{setspace}
\usepackage{ragged2e}
\usepackage{titlesec}
\usepackage{enumitem}
\usepackage{array}
\usepackage{longtable}
\usepackage{calc}
\usepackage{booktabs}
\usepackage{caption}
\usepackage{float}
\usepackage{xcolor}
\usepackage{etoolbox}
\usepackage[numbers,sort&compress]{natbib}
\setcitestyle{square}
\usepackage{hyperref}
\hypersetup{colorlinks=true,linkcolor=black,citecolor=black,urlcolor=black}

\title{Ontology-constrained multi-LLM scoring of hypothesis support in the predictive processing literature}
\author{Hamed Nejat$^{1,*}$, Alexander Maier$^{1}$, Jesse Spencer-Smith$^{2}$, Andr\'e M. Bastos$^{1}$}
\date{May 2026}

\justifying
\titleformat{\section}{\normalfont\bfseries\normalsize}{\thesection.}{0.5em}{}
\titleformat{\subsection}{\normalfont\bfseries\normalsize}{\thesubsection.}{0.5em}{}
\titleformat{\subsubsection}{\normalfont\bfseries\normalsize}{\thesubsubsection.}{0.5em}{}
\titlespacing*{\section}{0pt}{1.1em}{0.45em}
\titlespacing*{\subsection}{0pt}{0.9em}{0.35em}
\titlespacing*{\subsubsection}{0pt}{0.8em}{0.3em}
\setlist{itemsep=0pt,topsep=2pt,parsep=0pt,partopsep=0pt}
\AtBeginEnvironment{longtable}{\small\setlength{\tabcolsep}{3pt}}
\renewcommand{\arraystretch}{1.06}

\newcommand{\EquationBox}[3]{%
  \begin{center}
  \begingroup
  \small
  \setlength{\tabcolsep}{6pt}%
  \renewcommand{\arraystretch}{1.18}%
  \begin{tabular}{|p{0.27\linewidth}|p{0.64\linewidth}|}
  \hline
  \textbf{#1} & #2 \\
  \hline
  \multicolumn{2}{|p{0.94\linewidth}|}{#3} \\
  \hline
  \end{tabular}
  \endgroup
  \end{center}
}
\newcommand{\EquationBoxWideLeft}[3]{%
  \begin{center}
  \begingroup
  \small
  \setlength{\tabcolsep}{6pt}%
  \renewcommand{\arraystretch}{1.18}%
  \begin{tabular}{|p{0.24\linewidth}|p{0.67\linewidth}|}
  \hline
  \textbf{#1} & #2 \\
  \hline
  \multicolumn{2}{|p{0.94\linewidth}|}{#3} \\
  \hline
  \end{tabular}
  \endgroup
  \end{center}
}
\newcommand{\TemperatureEquationBox}{%
  \begin{center}
  \begingroup
  \footnotesize
  \fbox{%
    \begin{minipage}{0.86\linewidth}
    \textbf{Hypothesis space temperature}\hfill
    \(T_H=c\,V_{\mathrm{env}}/n,\quad c=1.0\)\par\vspace{0.35em}
    The scaling constant \(c\) is fixed to \(1.0\) for this analysis. Here, \(T_H\) denotes the metaphorical hypothesis-space temperature, \(V_{\mathrm{env}}\) is the volume of the minimum volume enclosing ellipsoid containing all hypothesis points, and \(n\) is the number of points. (eqn. 3.4)
    \end{minipage}%
  }
  \endgroup
  \end{center}
}

\makeatletter
\renewcommand{\maketitle}{%
  \begin{flushleft}
  {\large\bfseries \@title\par}
  \vspace{0.9em}
  {\normalsize\textbf{Authors:} Hamed Nejat$^{1,*}$, Alexander Maier$^{1}$, Jesse Spencer-Smith$^{2}$, Andr\'e M. Bastos$^{1}$\par}
  \vspace{0.9em}
  {\normalsize\bfseries Affiliations\par}
  \vspace{0.5em}
  \begin{enumerate}[leftmargin=0.75in,label=\arabic*.]
  \item Department of Psychology and Vanderbilt Brain Institute, Vanderbilt University
  \item Data Science Institute, Vanderbilt University
  \end{enumerate}
  \vspace{0.3em}
  {\normalsize $^*$Corresponding author.\par}
  \end{flushleft}
  \vspace{0.4em}
}
\makeatother

\begin{document}
\pagestyle{plain}
\maketitle

\noindent\textbf{Abstract}\par
Fragmentation is common in interdisciplinary fields with diverse methods and theoretical commitments. Predictive coding neuroscience is a clear example: its literature spans computational theory, electrophysiology, imaging, behavior, and modeling, creating a synthesis problem that conventional meta-analysis cannot easily resolve. Here, we describe a local multi-LLM pipeline for ontology-constrained literature synthesis. The pipeline reads papers, extracts evidence, incorporates figure descriptions, assembles constrained prompts, and validates outputs against an expert glossary. We manually defined a predictive-coding glossary of thirty-six concepts grouped into three hypotheses: predictive suppression, feedforward error propagation, and ubiquity. A council of ten local language models scored 31 studies according to their agreement or disagreement with each glossary factor across local and global oddball contexts. This enabled pairwise study-agreement analysis, cross-model comparison, and three-dimensional hypothesis-space mapping. Agreement was high for some hypotheses but weaker for others, revealing structured disagreement, particularly across local versus global oddball paradigms. We further define hypothesis-space temperature, a geometric dispersion metric measuring how compactly studies occupy the hypothesis space. Temperature was lower for local oddball contexts and higher for global oddball contexts, indicating greater dispersion in the latter. The scoring geometry also allowed us to estimate vectors of change between experimental contexts. These results demonstrate that local multi-LLM councils can produce auditable disagreement measurements that map heterogeneous literatures into quantitative evidence spaces. This framework may generalize to cross-study hypothesis mapping where conventional meta-analysis lacks a common comparison space.

\vspace{0.8em}

\section{Introduction}\label{introduction}

\subsection{The Fragmentation Problem in Predictive Coding Neuroscience}\label{the-fragmentation-problem-in-predictive-coding-neuroscience}

Scientific frameworks rarely develop through empirical support alone\cite{ref1}. They also face growing theoretical and methodological complexity, becoming large, specialized, and heterogeneous enough that readers may struggle to sustain the full argumentative structure\cite{ref2,ref3,ref4,ref5}. This ``fragmentation'' of the scientific literature is acute in the domain of computational neuroscience. In particular, mathematical formalism, electrophysiology, behavioral observation, statistical modeling, and philosophical foundations are frequently braided together within a single publication, yet the resulting constructs are not consistently tested at equivalent levels of description. The paradigm of predictive coding serves as an illustrative exemplar of this epistemological phenomenon. Over the past few decades, predictive coding has evolved from a narrow computational account of extra-classical receptive field effects in the visual cortex\cite{ref6} and the retina\cite{ref7} into a sweeping, nearly all-encompassing explanatory framework addressing perception\cite{ref8,ref9}, learning\cite{ref10}, motor action\cite{ref11,ref12}, cortical hierarchy\cite{ref13,ref14}, psychiatric disorders\cite{ref15,ref16}, and active sensing\cite{ref11,ref17}.

As the predictive coding framework has broadened its theoretical scope, the consensus on what constitutes rigorous empirical support has been put into question\cite{ref18,ref19}. At one end of the theoretical spectrum, predictive coding represents a formal, largely abstract principle of hierarchical inference: higher cortical areas issue generative predictions, lower areas compute sensory residuals or prediction errors, and these errors propagate forward up the cortical hierarchy to continuously update internal models\cite{ref20,ref21}. At another theoretical extreme, predictive coding is treated as a highly specific, falsifiable physiological hypothesis mapping directly onto canonical cortical layers, specific cell classes, distinct oscillatory bands, and directed laminar interactions\cite{ref14,ref22,ref23}. For instance, some physiological accounts explicitly associate feedforward prediction error signaling with superficial pyramidal populations communicating via high-frequency gamma rhythms\cite{ref24}, while feedback predictions are associated with deep pyramidal populations communicating via lower-frequency alpha and beta rhythms\cite{ref25,ref26}. Furthermore, the terminology itself is used somewhat inconsistently across the literature. Terms such as prediction, expectation, anticipation, surprise, mismatch, deviance, novelty, and error can be deployed in overlapping but distinct contexts depending on the subfield. For example, a reduction in neural activity upon repeated stimulus presentation might be categorized as prediction suppression, sensory adaptation, or attentional gating, depending on the context.

This ambiguity is magnified by the methodological diversity present in modern neuroscience. Predictive coding claims have been investigated using disparate tools ranging from event-related potentials\cite{ref27} and functional magnetic resonance imaging univariate contrasts\cite{ref28,ref29} to multivariate decoding\cite{ref30,ref31,ref32}, laminar probe recordings\cite{ref24,ref33,ref34}, calcium imaging\cite{ref35,ref36,ref37}, and a diversity of behavioral tasks\cite{ref8,ref24,ref38}, including studies using no-report paradigms\cite{ref34,ref36,ref39}. Because human experts read this literature through specialized methodological lenses, traditional meta-analytic synthesis across these subfields does not always scale and is difficult to audit. A case in point is a recent community experiment which performed a large-scale survey of the predictive processing literature\cite{ref40}. The result involved over one year of intellectual labor, weekly discussions amongst dozens of neuroscientists, and iterated writing and editing, which proceeded as shown in the flowchart of Fig. 1A. This process was valuable and produced consensus amongst the co-authors, while also illustrating the substantial coordination required for manual synthesis. The resulting manuscript was 141 pages long, providing an extensive qualitative synthesis rather than a compact quantitative map of the literature\cite{ref40}. Moreover, there is currently no widely established mechanism for how to incorporate an increasingly large number of new results into this and other established literature syntheses. We propose that this heterogeneity and the volume of scientific publication create a bottleneck for cumulative theory-building. In the following, we demonstrate an AI-assisted approach designed to address this problem.

\subsection{Integrating Large Language Models into Scientific Synthesis}\label{integrating-large-language-models-into-scientific-synthesis}

The growth of scientific publication volume motivates automated systems capable of reading, comparing, and structuring literature at large scales (Fig. 1B). AI models can supplement established paradigms for literature synthesis and quantitative meta-analysis, while preserving the value of expert interpretation and conventional meta-analytic methods.

\begin{center}
\includegraphics[width=0.95\linewidth,keepaspectratio]{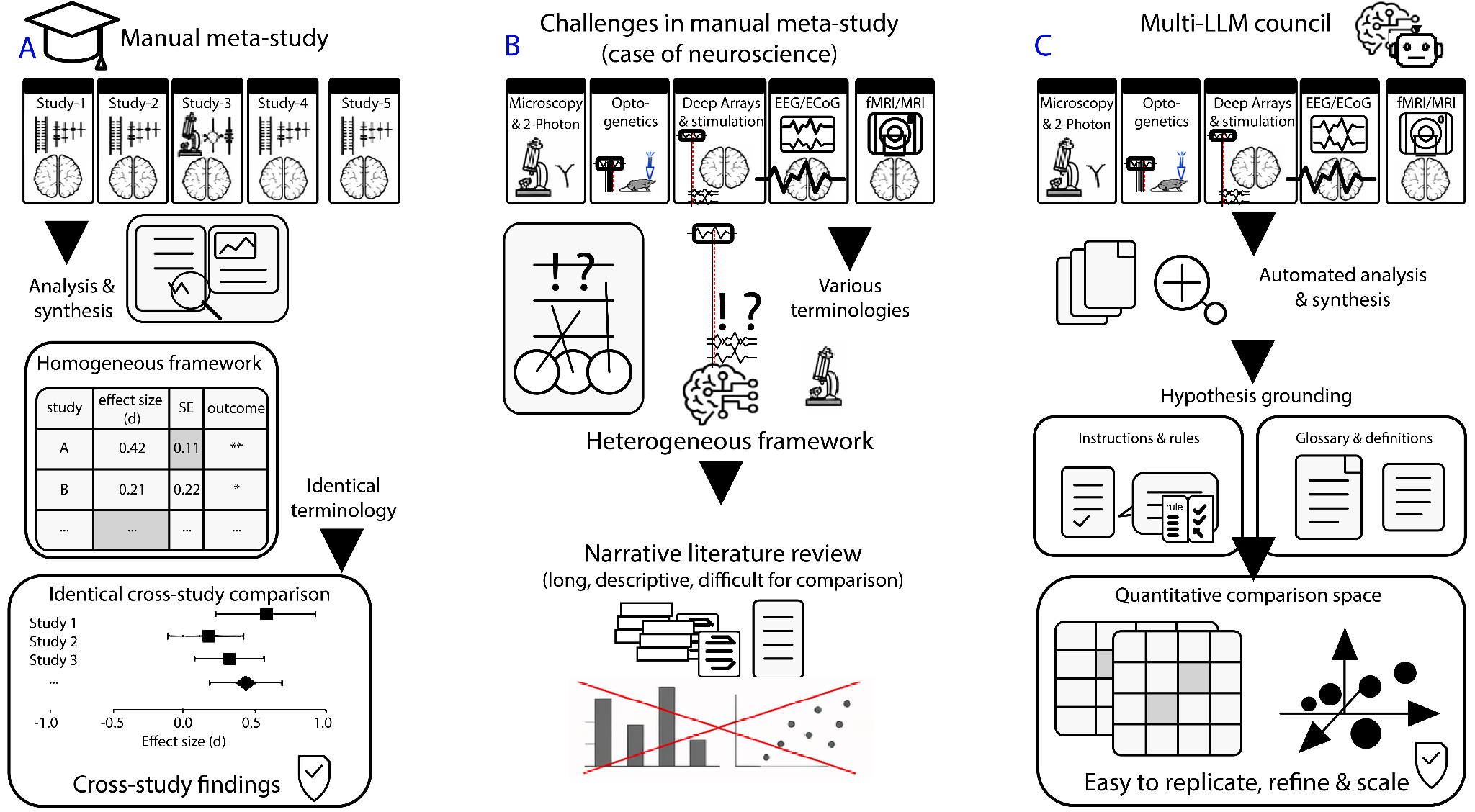}
\end{center}
\textbf{Figure 1:} A. Manual meta-studies (e.g., literature review, meta-analysis) are time-intensive and become harder to scale as the magnitude of the literature grows. B. The challenges with a traditional meta-analysis from the perspective of systems neuroscience. Heterogeneous data sources complicate unified statistical treatment. C. The LLM-assisted pipeline we propose here. AI-assisted literature review can ingest a large number of papers and score them on consistent criteria, which can be adjusted by experts. Multiple agents can be run in parallel in a scalable architecture. Replicable output scores can be quantified and compared between studies and agents.

First, in a traditional meta-analysis, papers are generally selected based on a literature which used the same techniques and the same paradigm, and therefore which exist within an identical comparison space (Figure 1A, for a recent application of this approach to the neurobiology of schizophrenia, see Mulvey, Gabhart, et al.\cite{ref41}). Under these constraints, each study's effect size and reliability is quantified and the consistency of the overall effect across studies is quantified with standard statistics. However, there are growing challenges to this traditional meta-analysis framework. This approach is difficult to apply to heterogeneous literatures where papers use different methodologies and experimental paradigms (Figure 1B). In such settings, there may be no established comparison space and no single standard statistical approach for quantitatively characterizing disparate results (Figure 1B). The common alternative is qualitative literature review, which is valuable but can be difficult to replicate and may not always compress information into comparable variables (for a recent case study in this, see\cite{ref40}).

To address these challenges, we propose an expert-guided AI-assisted pipeline (Figure 1C) similar to other recent work showing that LLMs can automate multiple aspects of systematic reviews\cite{ref42} and meta-analyses\cite{ref43}. The pipeline we propose can ingest and process a large number of heterogeneous papers in parallel and create a structured comparison space based on a canonical, expert-guided glossary. Based on this glossary, a council of multiple independent LLMs can quantitatively score papers on their agreement to a well-defined hypothesis space. Subsequently, the pipeline performs consensus mapping of papers within the hypothesis space. This effectively compresses a large amount of information from heterogeneous sources into a low dimensional structure of quantitative variables. The output is a geometric constellation, or object, that represents the scientific literature in relation to theoretical predictions that can be further analyzed. New papers can easily be added and compared. Movement of the constellation itself can be tracked over time as a literature evolves, reflecting growing agreement or disagreement with the hypothesis space.

To achieve these goals and to create the proposed pipeline (Figure 1C), it is important to evaluate recent progress of LLMs in scientific reasoning. Indeed, large language models have recently demonstrated extraordinary capabilities across diverse domains, yet their application in rigorous scientific synthesis remains an active, highly debated frontier\cite{ref5,ref44,ref45,ref46}.

Within the domain of literature synthesis a recent study (\emph{BrainBench}) showed that large language models tuned to neuroscience (e.g., \emph{BrainGPT}) can forecast neuroscientific results at a level rivaling or exceeding human domain experts\cite{ref5}. The \emph{BrainGPT} study demonstrated that by capturing the fundamental patterning of methods and results underlying the structure of neuroscience, predictive models can determine whether the scientific content of new abstracts were altered or not more reliably than humans at various levels of domain expertise\cite{ref5}.

Similarly, the emergence of autonomous research frameworks like \emph{The AI Scientist}\cite{ref45} have illustrated that multi-agent architectures can navigate the entire research lifecycle, from initial ideation and hypothesis generation to coding, result analysis, and even automated peer review (but see\cite{ref47}). The automated reviewer component of \emph{The AI Scientist}, which operates on an ensemble methodology to evaluate manuscript quality against established conference guidelines, proved capable of predicting paper acceptance decisions with an accuracy on par with human reviewers. The system also generated machine learning results that were accepted in a peer-reviewed conference\cite{ref45}.

More recently, several systems have moved toward structured scientific workflows. Empirical Research Assistance (ERA)\cite{ref48} coupled LLM-guided code rewriting with tree search to optimize empirical scientific software against explicit, machine-scored objectives, outperforming single LLM calls, best-of-1000 sampling, and AIDE on Kaggle benchmarks. Robin\cite{ref86} extended multi-agent systems into a lab-in-the-loop biological discovery setting by integrating literature-search agents with data-analysis agents to generate hypotheses, propose experiments, and interpret experimental results. Co-Scientist\cite{ref87}, similarly used an asynchronous multi-agent architecture, under expert supervision, to generate, critique, rank, and evolve scientific hypotheses. Together, these systems suggest that LLMs can contribute to scientific reasoning when constrained by external objectives, specialized agent roles, iterative feedback, and human or experimental validation. These systems also motivate a complementary problem: before autonomous systems generate or test new hypotheses, auditable methods are needed for mapping what the existing literature already supports, contradicts, or leaves unresolved. Here we propose that deploying a constrained, multi-model council equipped with expert-guided ontological guardrails offers a robust mechanism for parsing a fragmented scientific literature like predictive coding.

\subsection{Concrete Use Case: Oddball Paradigms in Predictive Coding}\label{concrete-use-case-oddball-paradigms-in-predictive-coding}

Here we examine a conceptually clean yet empirically nontrivial distinction within predictive processing: the separation of the local-oddball (LO) and global-oddball (GO) contexts, a paradigm which has been extensively studied and which has been considered as foundational evidence for predictive processing models in neuroscience\cite{ref49}. The local-global oddball paradigm was explicitly developed to differentiate short-timescale sensory regularity violations from longer-timescale, abstract rule, or sequence violations\cite{ref29}. In the local condition, sensory deviance is defined relative to an immediate, highly probable sensory context, such as a single deviant tone following a rapid succession of standard tones (e.g., xxxY, where ``x'' and ``Y'' represent distinct tone frequencies presented to subjects within a few hundred milliseconds). In the global condition, deviance is mapped against a higher-order temporal or abstract sequence expectation (e.g., xxxY xxxY xxxX), requiring the brain to maintain working memory of complex rules across extended temporal windows.

If the predictive coding framework operates as a unified, scale-invariant hierarchical architecture, the generative logic governing the local oddball response should extend to the global oddball response: Higher-level rule violations should generate prediction errors that propagate forward using the same basic inferential mechanics as local sensory violations\cite{ref49}. However, the empirical literature is still unclear on this point\cite{ref33,ref34,ref49,ref50}. Divergent neural signatures\cite{ref34}, varying attentional dependencies\cite{ref29}, unique conscious-state sensitivities\cite{ref33,ref51}, and differing spatial cortical distributions\cite{ref29,ref34,ref52} all complicate a direct one-to-one mapping between local and global contexts. This empirical tension makes the local-global paradigm a useful benchmark for evaluating scientific literature on predictive processing using LLM assistance.

\subsection{Rationale for a Local Multi-LLM Council Over a Single Model}\label{rationale-for-a-local-multi-llm-council-over-a-single-oracle}

A single large language model (LLM) can provide a useful literature synthesis, but a multi-model council offers several methodological advantages. First, model outputs can reflect model-specific priors and blind spots\cite{ref53,ref54}. Second, a single output provides limited information about internal disagreement. In contrast, deploying a benchmarking ``council''\cite{ref55} composed of multiple local language models makes disagreement measurable. Third, multiple agents can help identify unsupported assertions because pooling outputs across agents reduces the influence of uncorrelated model-specific errors\cite{ref46}. Fourth, multi-model ensembles can improve performance relative to a single model in benchmark settings such as medical question answering\cite{ref56}. Given these advantages, we used multiple LLM agents to independently score each paper.

Deploying local models provides additional practical and governance advantages. Local execution ensures that the entire analytical process can be accessed and audited. Intermediate markdown files, vision-language model figure extractions, prompt templates, validation statuses, and reasoning logs remain inspectable on local hardware. Furthermore, this approach avoids exposing proprietary scientific documents to external cloud application programming interfaces, ensuring strict data governance, traceability, and reproducibility with fixed review processes, which are paramount for the scientific integrity of automated literature reviews.

\begin{center}
\includegraphics[width=0.86\linewidth,keepaspectratio]{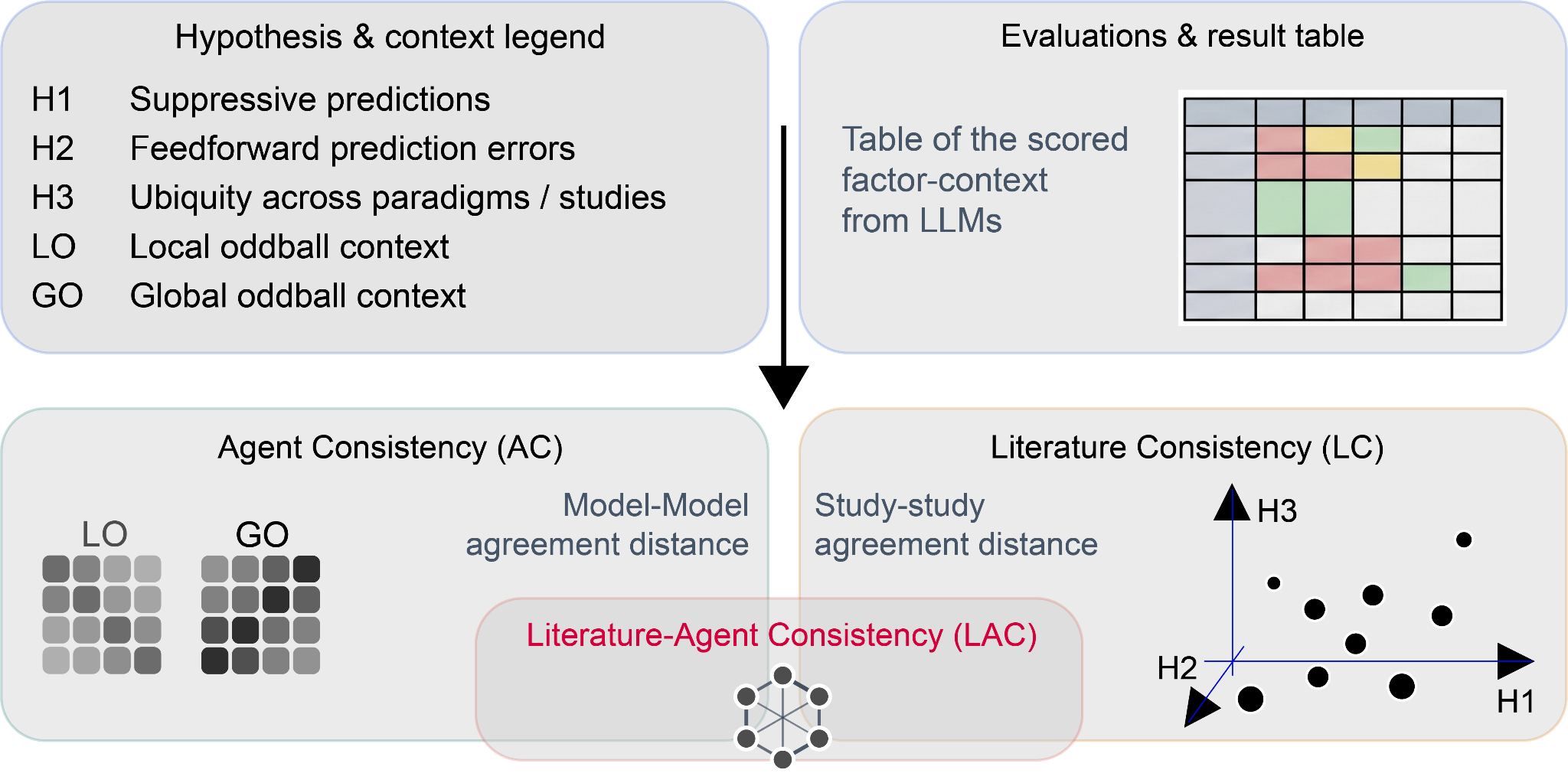}
\end{center}
\textbf{Figure 2: Benchmark overview.} A 36-factor glossary defined the Hypothesis space (see Supplementary Tables S1-S3). Two separate contexts are defined, local and global oddballs. The corpus of papers (see Supplementary Table S5) were scored for agreement / disagreement with each hypothesis and in both contexts. Each agent performed these evaluations and saved the structured output in a table. Quantitative analysis of this table defines three benchmarks: \emph{Agent Consistency} (AC, equation 1.1, 1.2) measures how similar different LLMs were to each other in their scores. \emph{Literature Consistency} (LC, equation 2.1, 2.2) measures how similar different papers were to each other in their scores on each of the three Hypotheses. The distance between each paper's position in this 3D space represents a metric of disagreement between any two sets of findings in the literature. \emph{Literature-Agent Consistency} (LAC, equations 3.1-3.3) quantifies a combination of across-agent and across-paper variability.

\section{Results}\label{results}

\subsection{Agreement Across Hypotheses and Contexts}\label{agreement-across-hypotheses-and-contexts}

The local language model council successfully transformed the heterogeneous predictive coding literature into a structured, measurable evidence space. The aggregated results shown in Figure 3 revealed that model evaluations are nonuniform and depend on both the hypotheses (Hypothesis 1, 2, and 3) and the context scope (local and global oddball). The models interpreted most papers as agreeing with these three hypotheses, which form a foundation of the predictive coding literature\cite{ref10,ref14,ref59}.

Within the local oddball context, the highest score was achieved by evaluations of Hypothesis 2 (Feedforward Error Propagation, Mean \(=0.51 \pm 0.34\), SD). The next highest score was Hypothesis 1 (Predictive Suppression, Mean \(=0.46 \pm 0.42\), SD), followed by Hypothesis 3 (Ubiquity, Mean \(=0.31 \pm 0.39\), SD). The convergence on H1 and H2 suggested that the evidence of suppression-like effects of prediction --- such as response dampening or repetition suppression --- were consistently described in the literature and had a feedforward pattern of activation in cortical brain areas. Furthermore, these consistently high scores were identified by the diverse models that comprise the council. In contrast, the ubiquity hypothesis (H3, across both contexts) displayed the lowest agreement (Mean \(=0.23 \pm 0.40\)), which can be interpreted to indicate that models identified less evidence (compared to H1 and H2, across both contexts) for these prediction error signals to generalize across the brain.

An assumption within hierarchical predictive coding framework is that global contextual prediction inherits the computational machinery of local sensory prediction, operating on similar predictive coding principles albeit at longer timescales\cite{ref14,ref69,ref80}. The global oddball context was also positive for H1-H3, but the scores were lower than the local oddball context for all three hypotheses. There was an average global-to-local downward shift of -0.049 for H1 (Predictive Suppression, p < 0.05), -0.158 for H2 (Feedforward Error Propagation, p < 0.01), and -0.032 for H3 (Ubiquity, p < 0.05, all comparisons, LO vs.~GO, paired model-study t-test). The lowest scores across all hypotheses/context groupings were observed in the global oddball context for the ubiquity hypothesis (H3). Across the thirty-one-study corpus, the council consistently recorded lower scores in global oddball contexts.
\begin{center}
\includegraphics[width=0.88\linewidth,keepaspectratio]{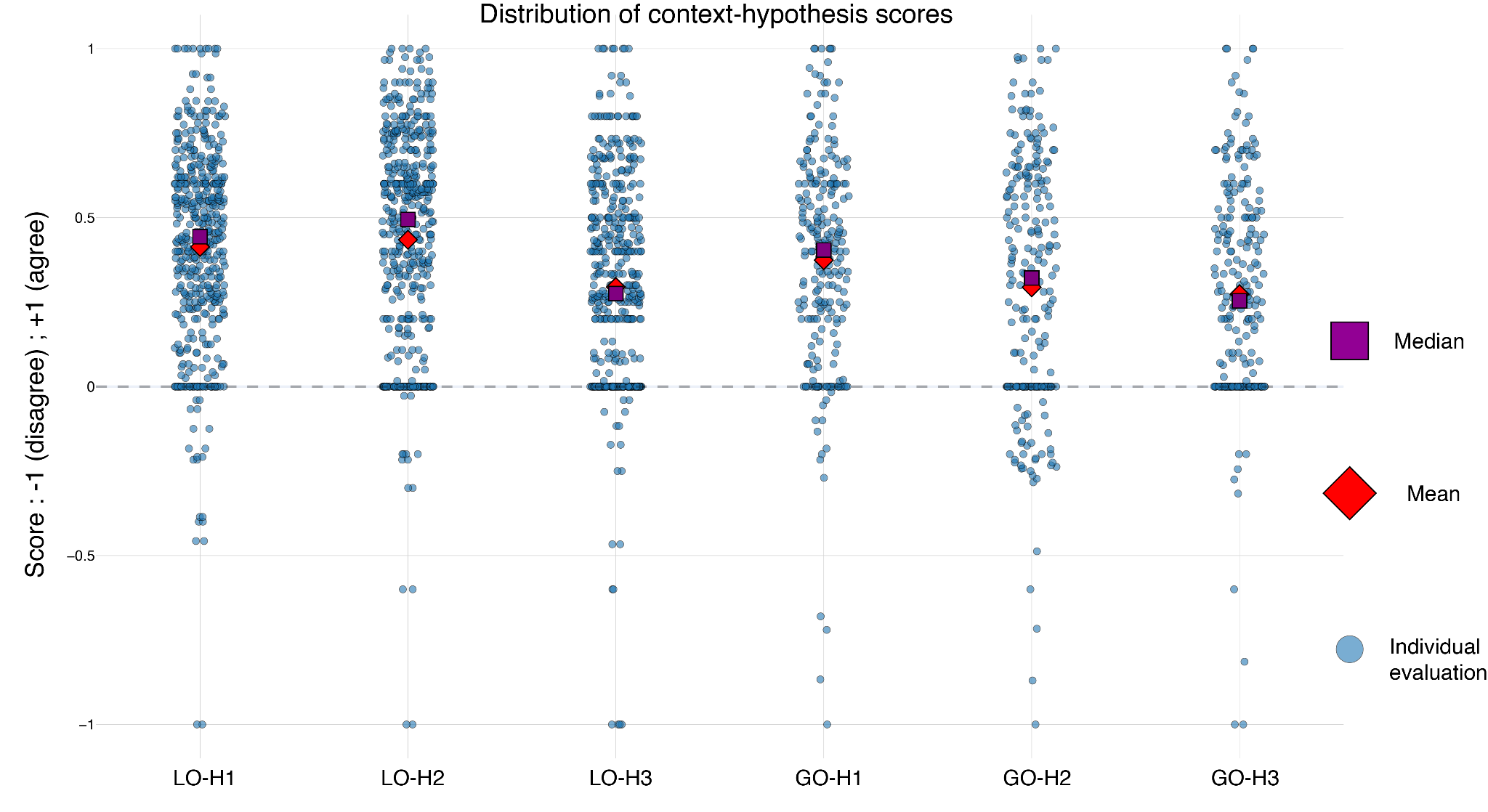}
\end{center}
\textbf{Figure 3:} The evaluation result of all papers (N=31) by all models (N=10). Scores are separated by context (LO = Local Oddball, GO = Global Oddball) and by Hypothesis (Supplementary Tables S1-S3, Hypothesis 1: Predictive Suppression; Hypothesis 2: Feedforward Error Propagation; Hypothesis 3: Ubiquity). Each dot is a separate score (total count = 1860). The mean and median of the scores are significantly positive (p < 0.01), which is consistent with our choice of these hypotheses as central to predictive coding.

To quantify inter-model agreement for both the local and global oddball context and separately for H1-H3, we used the \emph{Mean Squared Difference} as a distance metric (see Methods, Agent Consistency, equation 1.1, 1.2). Lower values indicate more similar values and therefore stronger agreement/consensus among the models. Agreement between models (Figure 4, AC) had a mean of 0.11. This was well below what could be expected from a random shuffle (0.67) or a hypothesis shuffle (0.5) (for the definition of these shuffle tests, see Methods section, definitions 1.8 and 1.9).

\begin{center}
\includegraphics[width=0.95\linewidth,keepaspectratio]{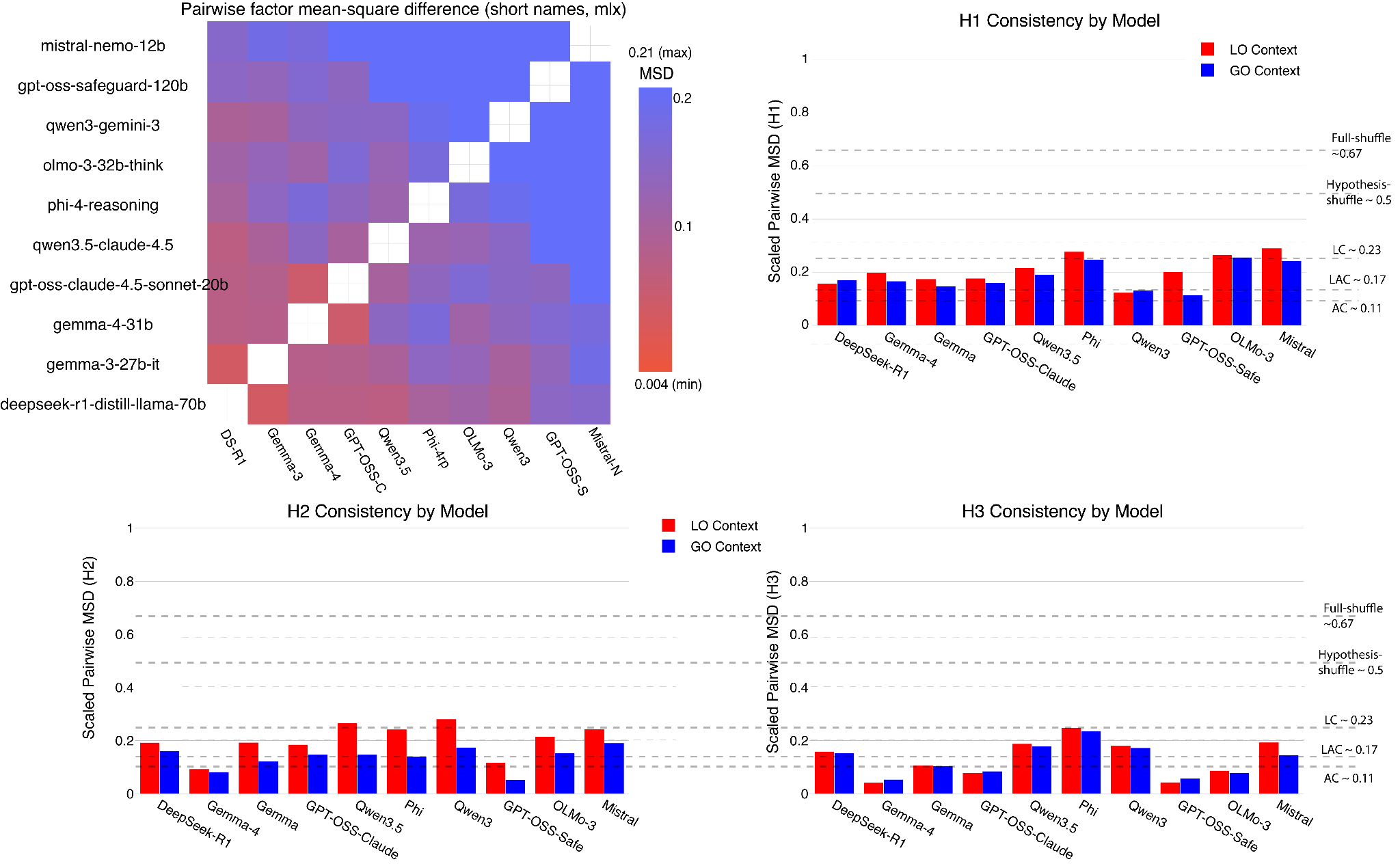}
\end{center}
\textbf{Figure 4:} Agent consistency measured as the mean squared distance between the aggregated scores of each of the 10 agents. Agent to agent distance is shown on the upper left. The mean in this comparison was 0.11. Mistral-nemo-12b was the model with the largest distance to the others. The mean squared distance of each model by context is shown for H1 (upper right), H2 (lower left), and H3 (lower right). The dotted lines indicate the MSD distance for a random shuffle of all scores (Full-shuffle, 0.67, def 1.8) and the Hypotheses shuffle (0.5, see Methods, def.1.9). For comparison, values of AC, LC, and LAC are also depicted (see Methods for their respective definitions).

The \emph{Literature Consistency (LC)} benchmark enabled us to quantitatively identify how variable scores were between the papers we considered (Figure 5). Studies such as Bastos et al., 2012, Friston et al., 2010, and Spratling, 2008 exhibited low mean distances (\(0.07 \pm 0.01\) SEM), marking them as central reference points that closely align with the dominant trends of the literature. Conversely, Westerberg, Xiong, et al., (2025) emerged as an outlier, with a mean squared difference of 0.464 relative to the rest of the literature. This is consistent with the formulation and conclusions of Westerberg, Xiong, et al., (2025), which diverge from several predictive coding hypotheses, especially in the global oddball context. The LC benchmark therefore enables a geometric mapping between papers and identifies publications that are central to, or divergent from, the established hypotheses.

\begin{center}
\includegraphics[width=0.92\linewidth,keepaspectratio]{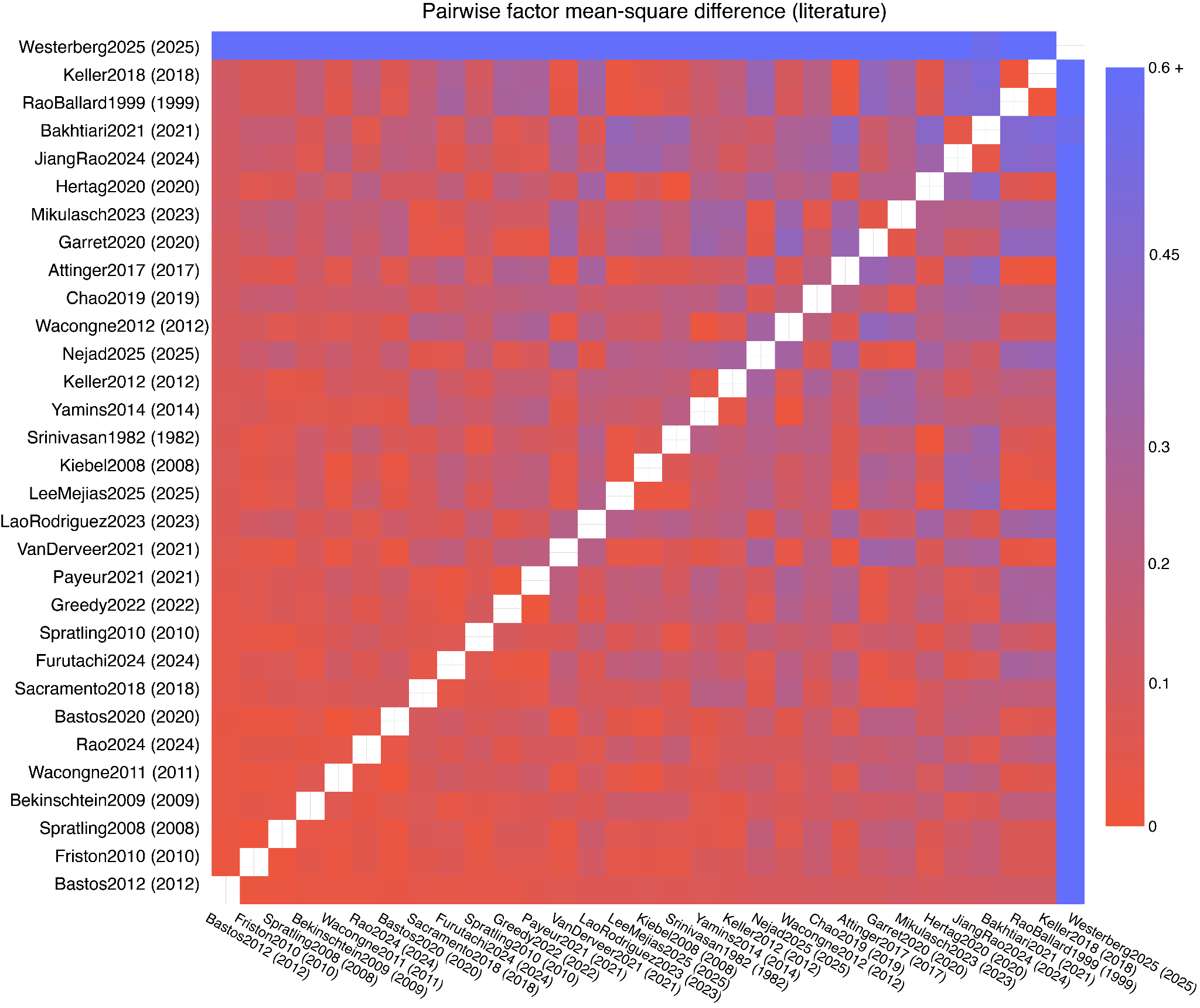}
\end{center}
\textbf{Figure 5:} Literature consistency (LC) visualization. The mean squared distance between every paper in the corpus is depicted. Warmer colors indicate lower MSD values or more agreement between studies. The highest values were observed for an outlier paper, Westerberg et al., 2025.

\subsection{Direct Comparison of Local and Global Contexts between papers}\label{direct-comparison-of-local-and-global-contexts-between-papers}

Further analysis of the distance scores between Westerberg, Xiong, et al., (2025) and the remaining literature revealed an interesting pattern: the scores were significantly greater for the global oddball (p < 0.01, mean = 0.55) as compared to the local oddball context (mean = 0.15), see Figure 6. This indicates that the council benchmark captured the pattern reported by Westerberg et al.\cite{ref34}: stronger divergence from predictive-coding expectations in global oddball findings than in local oddball findings.

\begin{center}
\includegraphics[width=0.95\linewidth,keepaspectratio]{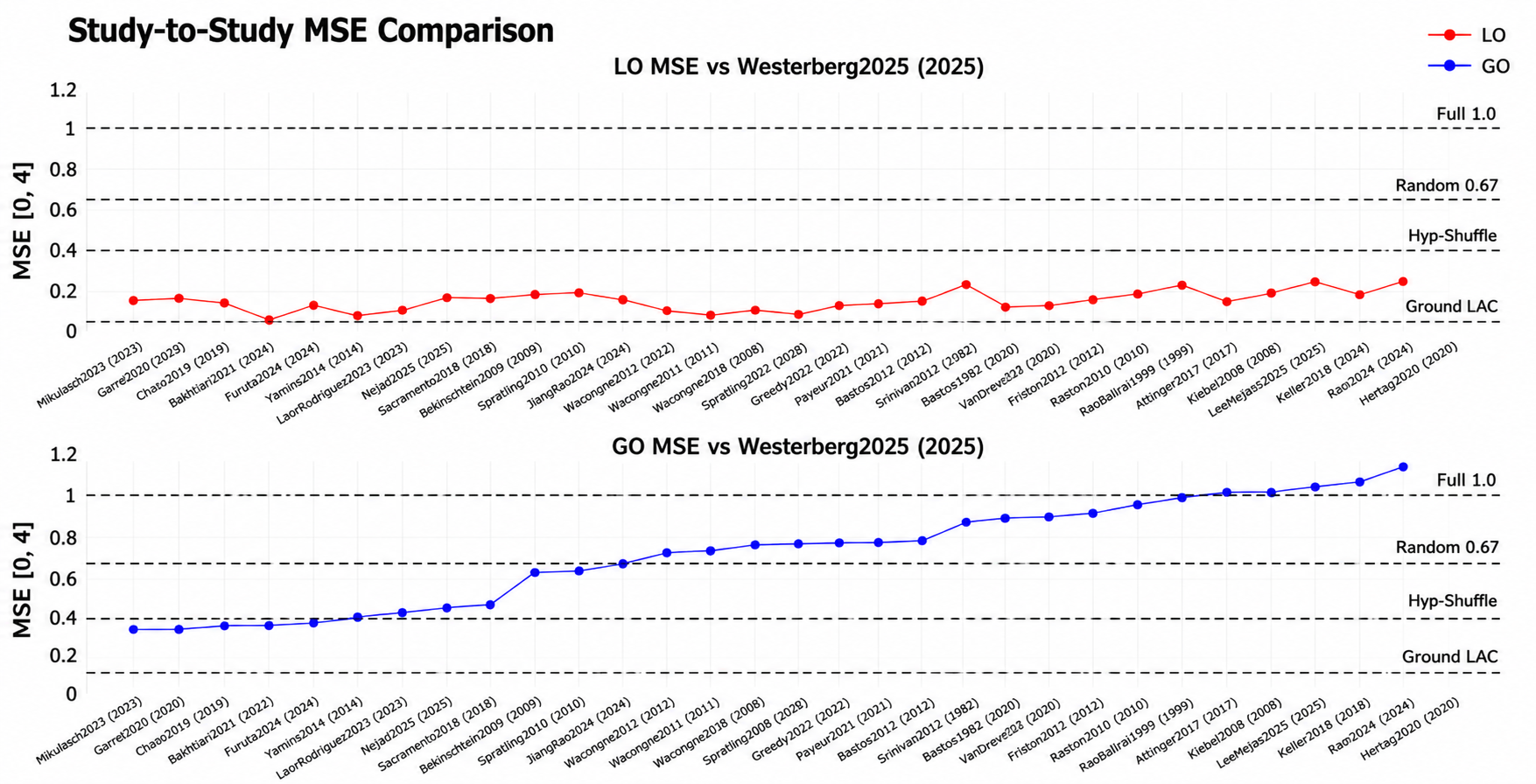}
\end{center}
\textbf{Figure 6:} Mean squared error values measuring the distance between each paper and the outlier paper (Westerberg, Xiong, et al., 2025), separately for the local oddball (upper subpanel) and global oddball (lower subpanel) contexts. Papers are ordered from left to right in order of increasing global-oddball MSD to Westerberg, Xiong, et al.~(2025).

\subsection{Three Dimensional Hypothesis Space Geometry Reveals a Constellation}\label{three-dimensional-hypothesis-space-geometry-reveals-a-constellation}

To visualize the latent structure of agreement and disagreement within the literature, the studies were geometrically embedded into a three-dimensional hypothesis space defined by their H1, H2, and H3 scores, when rendering the local and global contexts separately. The central coordinate of each study reflects the average score separately per hypothesis and averaged across the council of 10 models. Error bars denote the standard error of the mean across the council. Within the local oddball coordinate space, the majority of the studies cluster into a moderately compact, centralized region along the H1, H2 and H3 axes, visually representing the foundational agreement the council reached regarding overall agreement on these hypotheses. For local oddballs, the one exception was Westerberg et al.~(2025), which had positive scores on H2 and H3, but a moderately negative score on H1 (Predictive Suppression).

\begin{center}
\includegraphics[width=0.94\linewidth,keepaspectratio]{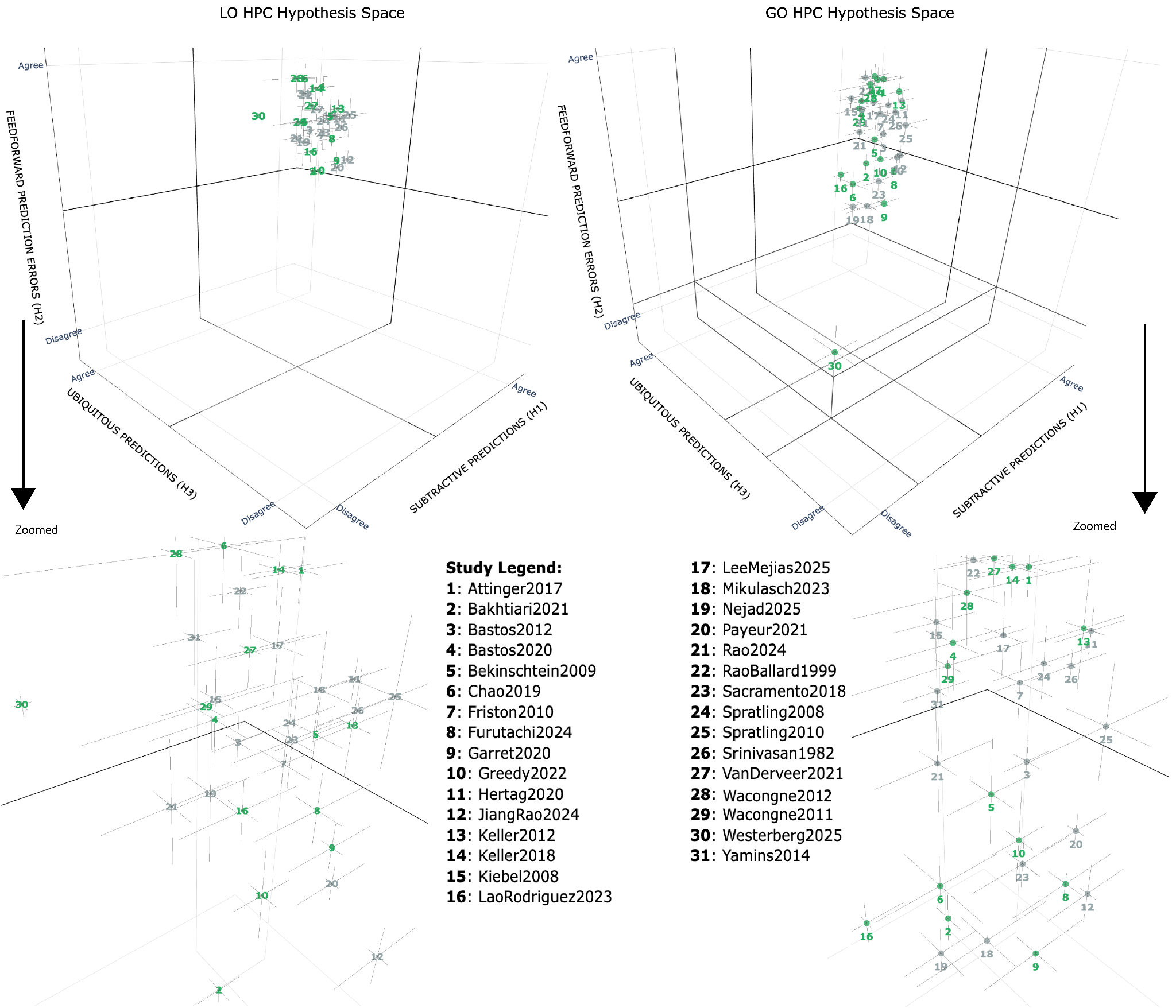}
\end{center}
\textbf{Figure 7:} Three-Dimensional view of the scores of each paper, with scores shown on the H1 (Predictive Suppression), H2 (Feedforward Prediction Errors), and H3 (Ubiquity) axes. The 3 lines emanating from each of the studies reflects the standard error of the mean of those scores across the 10 evaluator models. The local oddball context is shown on the left plot and the global oddball context is shown on the right.

Moving from local oddball (Figure 7, left) to global oddball (Figure 7, right) there was a significant move away from the +1, +1, +1 octant, with multiple studies going negative on H2. Study 30 (Westerberg, Xiong, et al., 2025) is the outlier, as it had the strongest movement between Local and Global Oddball contexts (Figures 7 and 8). The move from Local to Global Oddball contexts created a literature that was more dispersed or variable.

To quantify this shift, we introduced a novel metric of dispersion or variability within a well-specified hypothesis space. Inspired by physics, we call this metric the hypothesis-space-temperature. We define the hypothesis-space-temperature by fitting an ellipsoid that minimizes the volume required to fit all points in three dimensions, divided by the number of data points (studies) that are inside of the ellipsoid (see equation 3.4):

\TemperatureEquationBox
Operationally, when a hypothesis space provides a compact description of a literature, a new observation is expected to fall within the existing ellipsoid. In that case, the ellipsoid volume can remain approximately stable while the number of samples increases, thereby decreasing temperature. In contrast, a new study that falls outside the existing ellipsoid can increase the ellipsoid volume, increasing temperature in proportion to \(V/n\) (see equation above).

We calculated hypothesis-space-temperature separately for the Local and Global oddball contexts, and extracted the standard error of the mean of this measure across different LLM models. The hypothesis-space-temperature of Local Oddballs (excluding Westerberg, Xiong et al., 2025) was 0.00114 (SEM across N=10 models, 0.00016, p < 0.01), significantly lower than for Global Oddballs, 0.00348 (SEM across N=10 models, 0.00039, p < 0.01). This confirmed what can be appreciated visually in Figure 7, namely that the dispersion of studies grows larger in the Global Oddball compared to Local Oddball contexts, even without the outlier study. This effect was even stronger when including all 31 studies: across all studies, hypothesis-space-temperature of Local Oddballs was 0.00177 (SEM across N=10 models, 0.00033), significantly lower for Global Oddballs, 0.00566 (SEM across N=10 models, 0.00111). This geometric visualization indicates that the predictive coding literature is structured by a dense central neighborhood and distinct, isolated data points that increase dispersion in the hypothesis space.

\begin{center}
\includegraphics[width=0.86\linewidth,keepaspectratio]{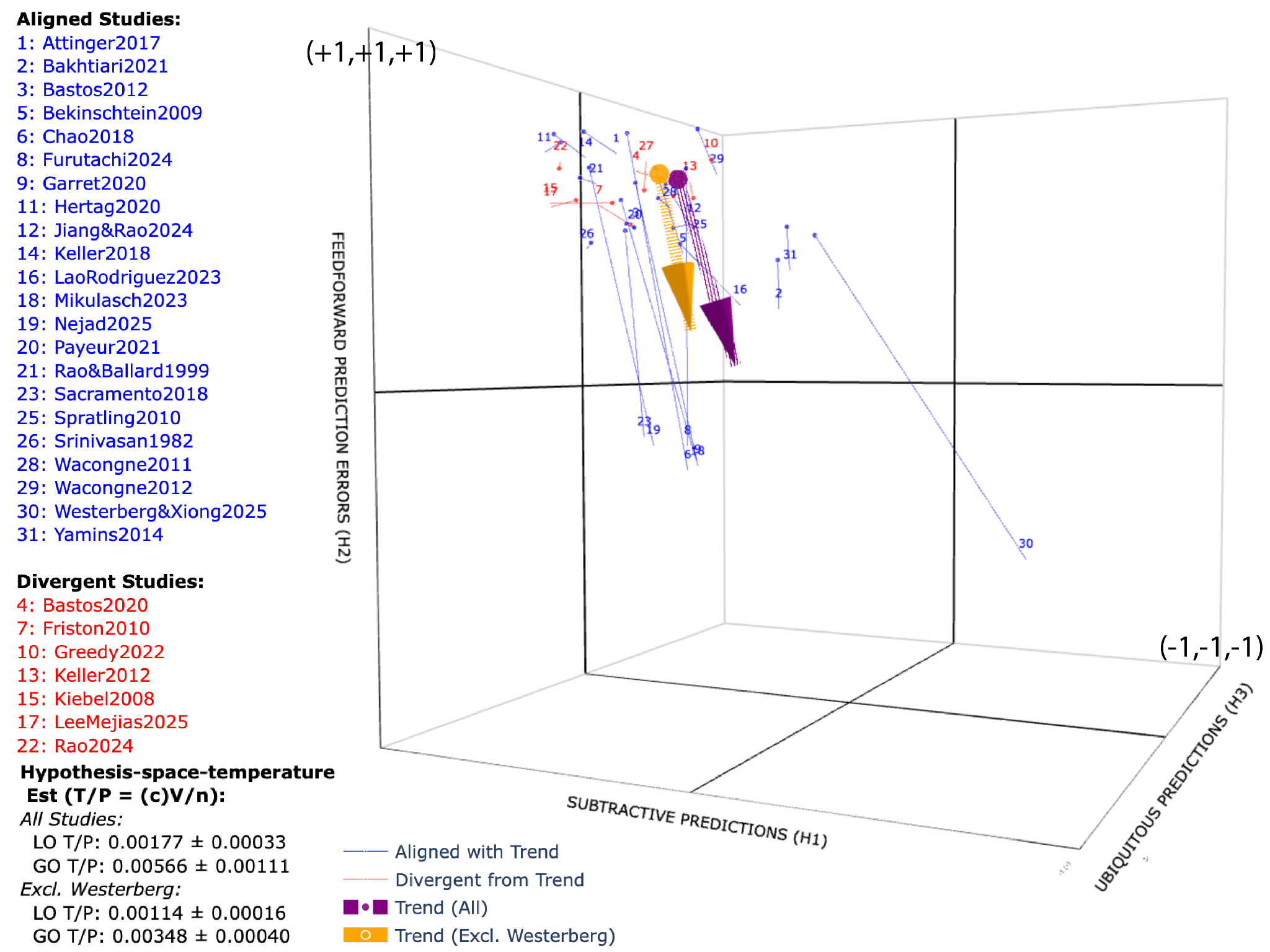}
\end{center}
\textbf{Figure 8:} Three-Dimensional vectors of each paper from the Local to the Global oddball position. Each paper's vector is shown independently. The average movement (vector direction) without the outlier study (Westerberg, Xiong, et al., 2025) is shown in yellow, and the average movement of all studies is shown in purple. Individual study's vectors are labeled as blue if their individual vector was aligned to the Westerberg, Xiong, et al., (2025) study's vector on all three axes, and red otherwise.

In addition to an increase in temperature, it is also apparent in Figure 7 that the change from Local to Global Oddball contexts appears to shift along a particular axis. To quantify this, we calculated a vector for each study between its Local and Global Oddball position. This is shown in Figure 8. Each vector represents an individual study's movement magnitude and direction. The mean vector across all studies is depicted in purple, while the vector of all studies excluding Westerberg, Xiong et al.~(2025) is shown in yellow. Both vectors are aligned in a similar direction pointing away from the +1, +1, +1 octant (which would indicate full agreement with all three hypotheses). We interpret this to mean that the literature largely agrees with the predictive coding hypothesis space for Local Oddballs but much less agreement is observed for Global Oddballs. Furthermore, while the Westerberg, Xiong, et al.~study is an outlier in the sense that the movement vector magnitude is the largest of all studies, the movement vector angle of this study is aligned to the underlying Local-Global shift within the pre-existing literature.

\section{Discussion}\label{discussion}

\noindent\textbf{Hidden structure within the predictive coding literature.}

The application of the multi-model benchmarking council shows that a heterogeneous predictive coding literature can be organized into structured, quantitatively identifiable patterns. The council's outputs showed that the literature can be organized by structured, quantitatively identifiable scores that draw out important distinctions between papers and contexts. One pronounced distinction was identified between stronger local oddball support and lower model-estimated support in the global oddball context. The models' agreement scores for H1, H2, and H3 were all higher for the local as compared to global contexts, which indicates stronger evidence for predictive coding in the local oddball domain. This context-dependent reduction was most pronounced for H2 (Feedforward Error Propagation), indicating that support for this hypothesis varies most strongly between local and global oddball contexts.

Furthermore, formalizing this quantitative hypothesis space permitted the extraction of geometric parameters of the corpus. Mapping the change of each individual study's movement from local to global paradigm with vectors allowed us to calculate theoretical shifts via mean vector displacement. This metric models directional movement between local and global paradigms, highlighting differences in these experimental contexts. Also, we derived a hypothesis-space-temperature metric. High temperature indicates hypothesis state-space volatility and theoretical disagreement. Low temperature signifies structural consensus. This provides the compass for a quantitative cartography of the literature, translating abstract scientific debate into measurable variables.

\noindent\textbf{Scientific synthesis with multi-model councils.}

For neuroscience and related interdisciplinary fields, our findings support the utility of multi-agent architectures for complex scientific analyses, echoing other AI-based advances such as \emph{The AI Scientist}\cite{ref45}. A single language model can summarize these thirty-one papers, but a council provides additional visibility into model-to-model variability, theoretical disagreement, and methodological nuance present in the corpus.

In contrast, the council framework quantifies and utilizes disagreement as an analytical tool. By calculating \emph{Agent Consistency}, \emph{Literature Consistency}, and \emph{Literature-Agent Consistency} baselines, researchers can map whether a particular scientific result is stable across contexts or context-sensitive and model-specific. In addition, random shuffling of model-produced values defines a statistical reference distribution under random scores. Finally, multiple model scores allow uncertainty over models to be estimated; here we used the standard error of the mean (Figure 7). The council framework therefore treats language-model outputs as structured, auditable scientific measurements when context and instructions are constrained.

\noindent\textbf{Importance of an explicit glossary and ontology.}

A conclusion drawn from this study is that, in artificial-intelligence-assisted scientific synthesis, the ontology matters alongside the underlying foundation models. The glossary provides the council with a stable conceptual target. The instruction layer and anti-guess rules\cite{ref58} further constrain model outputs toward explicit textual evidence rather than model-specific priors. In addition, all scores were accompanied by an auditable reasoning log. Inspection of a subset of these logs (see Section 2.4) revealed domain-appropriate reasoning patterns, while also identifying occasional divergences. These considerations suggest that the quality of language model benchmarking depends on task design, problem framing via a glossary, and the deployment of control layers. The ontology provided by the canonical glossary instructs the models what to look for, and allows us to understand and interpret what the benchmark signifies.

\noindent\textbf{Methodological limitations and future expansions.}

This framework has several methodological boundaries that guide future expansion. The corpus size of thirty-one studies, drawn largely from a recent study's comparisons\cite{ref34}, supports a focused initial cartography rather than exhaustive coverage of the predictive coding field. Second, the pipeline relies on the specific thirty-six factors chosen. Designing an alternative ontology based on active inference or variants of predictive coding theories could shift the geometric clustering. Because scores are nested within papers, models, factors, and contexts, future extensions can use hierarchical or mixed-effects models to estimate uncertainty at each level.

Future iterations of the pipeline can expand the corpus to enable higher-resolution analyses across decades of publications and additional subfields. In principle, a seed study could initiate corpus expansion through automated citation and similarity search. Future work can also examine whether ontological definitions of the hypothesis space can be proposed or refined automatically. In that setting, more than three dimensions could be included, and hypothesis dimensions could be added or revised as the literature evolves. This also points to the possibility of measuring literature dynamics over time. As scientific literatures shift, new findings revise established interpretations, and support for hypotheses waxes or wanes, extensions of the present approach could capture and quantify these changes, allowing meta-scientific analyses to guide future studies.

\noindent\textbf{Conclusions.} This research establishes a local multi-LLM benchmark framework designed to evaluate the predictive coding literature in neuroscience. By using ten local language models with a grounded glossary, instruction layers, visual extraction, and validation, the pipeline converts heterogeneous scientific papers into a quantifiable evidence space. Within this framework, assistance from large language models can be effective at mapping the empirical support for specific hypotheses that underlie scientific knowledge. The results illustrate that the literature supports consensus regarding local predictive suppression, but exhibits more disagreement concerning broad ubiquity claims and global feedforward error propagation. The integration of metrics such as \emph{Agent Consistency}, \emph{Literature Consistency}, and the \emph{Literature-Agent Consistency} provides a foundation for identifying consensus and empirical outliers. This enables expert-guided, automated, and replicable data-model comparisons between sets of results, including theoretical, conceptual, empirical, and computational papers. The distance metrics developed here provide a generalizable framework for hypothesis-space comparison. New evidence that differs from existing literature will be distant from the rest. As the weight of this evidence builds with additional datasets and published studies, the constellation of the underlying literature is altered. The methods introduced here can visualize and quantify these shifts. This captures the fundamental process of cumulative scientific reasoning itself, as additional evidence either fortifies existing hypotheses or renders them less likely. Additional hypotheses can also alter the geometric structure of scientific reasoning by adding additional dimensions beyond the three hypothesis space representations we considered here.

\section{Methods}\label{methods}

The multi-LLM processing pipeline, spanning from the initial ingestion of the raw publications to the visualization of the hypothesis space, was constructed to ensure that structured evidence can be traced directly back to its originating paper, its specific glossary factor, and its model-specific reasoning log.

\subsection{Overview of the Ontological Benchmark Pipeline}\label{overview-of-the-ontological-benchmark-pipeline}

The workflow was initiated by defining the hierarchical predictive coding framework and building the canonical glossary (Supplementary Tables S1-S3) and instruction layer. Following this, source PDFs (Supplementary Table S5) were ingested through deterministic text parsing. Each of a paper's figures were also ingested via a figure-aware step using \emph{DeepRead}\cite{ref57}. Once the study text and figure descriptions were consolidated into a unified markdown file, the system assembled a constrained evaluation prompt and dispatched it to the ten independent scoring models (Supplementary Table S6). The resulting outputs were validated and organized (based on the rules section of the input to the LLM) into a comprehensive table, which served as the database for all summary matrices and hypothesis-space visualizations (Supplementary Fig. S1).

\subsection{Corpus Structure and Benchmark Matrix}\label{corpus-structure-and-benchmark-matrix}

The working corpus (Supplementary Table S5) evaluated in the current study consisted of thirty-one published, preprint, and archived neuroscience studies, with publication years ranging from 1982 to 2025. These studies were selected because they span the heterogeneity of the predictive coding field, and included theoretical contributions, conceptual reviews, neurophysiology, functional imaging, and computational modeling studies. These studies were included in a recent paper as representative of predictive coding models\cite{ref34}. We recognize that they do not comprehensively represent the entirety of the predictive coding field, but are deemed sufficiently representative to demonstrate the utility of this pipeline. We plan to apply future extensions of this pipeline to a more comprehensive collection of papers, and the present work serves as a starting point. Performing these evaluations on a smaller set of papers also enables manual checking and verification of the results.

\subsection{Hierarchical Predictive Coding Glossary}\label{hierarchical-predictive-coding-glossary}

At the core of the benchmark is the thirty-six-factor glossary. Without a highly constrained glossary, large language models can rely on broad autoregressive priors, generating narrative summaries that may blur critical neuroscientific distinctions and the boundaries required for hypothesis testing\cite{ref58}. Following the logic proposed by Westerberg et al.\cite{ref34}, we defined a glossary organized by three distinct hypotheses: predictive suppression (H1), feedforward error propagation (H2), and ubiquity (H3). The full glossary, detailing the specific factors, their definitions, and their applicable context scopes, is provided in Supplementary Tables S1-S3.

\textbf{Hypothesis 1 (Predictive Suppression)} encompasses the core phenomenon that initially motivated predictive coding: that expected inputs are suppressed, sharpened, or explained away. Factors within this family (Supplementary Table S1), such as activity suppression or GABAergic inhibition serve to encompass the idea that top-down predictions suppress bottom-up sensory signals\cite{ref6,ref14}.

\textbf{Hypothesis 2 (Feedforward Error Propagation)} specifies that unpredicted mismatch is routed and propagated forward through the cortical hierarchy. This requires the models to search for signatures such as supragranular layer involvement\cite{ref59}, gamma-band signaling\cite{ref14}, and directed bottom-up connectivity\cite{ref24} (Supplementary Table S2).

\textbf{Hypothesis 3 (Ubiquity)} captures one of the most expansive theoretical claims, evaluating whether predictive coding serves as a universal algorithm of the cortex\cite{ref10,ref14}. H3 guides the models to discern whether a study provides evidence for mechanisms that are broadly consistent across-species, modalities, and/or brain areas (Supplementary Table S3).

Each factor was further described with more context regarding quantitative, qualitative, or methodological tags. The glossary dictated that not all textual evidence should be treated equally. A quantitative tag demands that the language model locate explicit statistical results, tables, or graphical data. The context scope column dictates the applicability of the factor to the local and global contexts. Factors directly tied to short-timescale adaptation effects are restricted strictly to local oddball scoring, preventing the benchmark from scoring sensory habituation as evidence for long-timescale sequence learning. The following papers were passed to the pipeline (Supplementary Table S5).

\subsection{The Instruction Layer, Prompt Methodology, and Constraints}\label{the-instruction-layer-prompt-methodology-and-constraints}

A glossary alone is likely insufficient if the language model acts as an unconstrained generative engine\cite{ref58}. Large language models can follow conversational priors, infer missing information, or substitute generic scientific plausibility for explicit textual evidence\cite{ref82}. Consequently, the pipeline separated the glossary from an instruction layer that specifies how the model evaluates the glossary.

The prompt architecture is initiated by establishing a specific role, instructing the language model to evaluate the paper as a neuroscientist. This role specification was intended to activate domain-relevant representations\cite{ref58}, orienting the model toward complex methodological details and the analytic standards typical of high-level peer review. The prompt then established a continuous scoring methodology for each factor ranging from negative one to positive one (within one decimal point, e.g., -0.2, +0.1). Positive scores indicate empirical support, negative scores indicate explicit contradiction, and precision is encouraged through the use of intermediate values to capture the magnitude of agreement/disagreement, given the evidence.

The prompt emphasized a distinction between scoring a zero and scoring a null value. A score of zero is assigned for instances where the study explicitly addresses a particular factor, but reports statistically neutral results that neither agree nor disagree with a factor. A null value was returned if the study does not address or measure the factor in any capacity (e.g., a study using EEG would not evaluate hypotheses related to cortical layers or specific cell types). Instructing the models to use null for unaddressed factors reduces false negatives and preserves the absence of evidence within the literature.

The instruction layer is further specified by anti-guess constraints. Models were instructed not to invent novel glossary keys, rename existing factors, or force local-global scoring onto irrelevant papers. Furthermore, the instructions required a reasoning log within the output. For every non-null score assigned, the model was required to cite the specific section, figure, or statistical table from the source text that justified its evaluation. This transformed the output from a standalone numerical assertion into an auditable chain of evidence, which we used to inspect whether unsupported assertions were present. We randomly inspected a subset of logs to confirm this. The full outputs are available on the repository (see \emph{Data and Code Availability}).

To give one instructive example, Mistral-Nemo scored Westerberg et al., 2025, with a reasoning that included ``\emph{\ldots{} I score H2: Feedforward Deviance as positive. In LO (Local-Oddball) Fig 2b, I find hierarchical progression {[}+0.8{]}. However, GO (Global-Oddball) signals in Fig 3 show no temporal order of response {[}-0.6{]}, suggesting an alternative mechanism compared to LO \ldots{}}''.

\subsection{Deterministic Literature Ingestion}\label{deterministic-literature-ingestion}

Text extraction alone is insufficient for neuroscience literature, where substantial inferential signal is frequently encoded within a paper's figures. A purely text-based evaluation pipeline would omit this empirical evidence. To address this, the pipeline incorporated a multimodal processing stage. The pipeline then extracted the figures and analyzed them with a local vision-language model (VLM) instructed to operationalize graphical data (gemma-3-9b-it VLM), decode statistical trends, and map visual evidence directly onto the predictive coding glossary. These descriptive insights were subsequently interleaved into the study markdown document which was passed to the LLM models.

\subsection{Model Council Orchestration and Validation}\label{model-council-orchestration-and-validation}

The following 10 models comprised the council of LLMs that were used for evaluation (Supplementary Table S6):

The pipeline contained a validation tool designed to audit syntactic integrity, verify top-level key presence, and ensure that factor nomenclature matches the canonical glossary. Any output that did not meet these constraints was flagged for exclusion or revision, preventing misaligned data (e.g., incorrect factor assignment) from entering the downstream hypothesis-space visualizations.

The LLMs in the pipeline were configured during model initialization via mlx-lm package utilizing the same sampling and memory optimization parameters for all models. Inference parameters were set to a sampling temperature\cite{ref84} of $\tau = 0.70$ with top-p/min-p of $p = 0.9$ and $\alpha = 0.1$, respectively. To manage computational overhead during long-context execution, the sequence length for each model was strictly bounded by a context window of $L = 2^{17}$ (131,072 tokens), with past attention states preserved in a KV-cache using MXFP8\cite{ref85} precision to optimize memory footprint and maximize throughput.

\subsection{Construction of the Hierarchical Predictive Coding Table and Measures of Evaluation Consistency}\label{construction-of-the-hierarchical-predictive-coding-table-and-measures-of-evaluation-consistency}

Validated outputs were flattened to construct the Hierarchical Predictive Coding Table. This matrix aligned the data into three hundred and ten rows, with each row corresponding to a unique study(s)-model(m) pair. The matrix comprises columns capturing metadata, and factor-level columns representing all thirty-six factors (f) across both local and global contexts (c). The table enables multidimensional hypothesis visualization and pairwise agreement correlation on the scores. Evaluation scores that were not applicable for a subset of factors were ignored and omitted from calculations.

The table holds values for each study s, model m, context c, and factor f, (def 1.1):
\EquationBox{Score Array Element}{\(S(s,m,c,f) \in [-1,1]\cup\{\emptyset\}\)}{Score value ranging from \(-1\) (disagrees) to \(1\) (agrees), or undefined (\(\emptyset\)). (def. 1.1)}
Where each of these s, m, c, and f elements contain the following indices (def 1.2):
\EquationBox{Score Indices}{\(s\in\{1,\ldots,31\},\; m\in\{1,\ldots,10\},\; c\in\{\mathrm{LO},\mathrm{GO}\},\; f\in\{1,\ldots,36\}\)}{Indices for study \((s)\), model \((m)\), context \((c)\), and factor \((f)\). (def. 1.2)}
The 36 factors are defined in Supplementary Tables S1-S3 and their indices correspond to Hypothesis 1 (Predictive Suppression), Hypothesis 2 (Feedforward Error Propagation), and Hypothesis 3 (Ubiquity). All evaluations had scores for each hypothesis to provide a non-missing numeric value for agreement with hypotheses, (def 1.3):
\EquationBox{Hypothesis Sets}{\(H_1=\{1,\ldots,12\},\; H_2=\{13,\ldots,24\},\; H_3=\{25,\ldots,36\}\)}{Partition of factors into three hypotheses. (def. 1.3)}
By averaging across studies (s), models (m), and contexts (c) it is possible to define each score for each hypothesis. Note that in the equation below, null values, which models are instructed to return when it is not possible to evaluate a factor, are removed from the mean (def 1.4):
\EquationBox{Mean Hypothesis Score}{\(\displaystyle \bar{S}_h(s,m,c)=\frac{\sum_{f\in H_h}\mathbf{1}[S(s,m,c,f)\ne\emptyset]S(s,m,c,f)}{\sum_{f\in H_h}\mathbf{1}[S(s,m,c,f)\ne\emptyset]}\)}{Mean score across factors in hypothesis \(H_h\) for fixed \((s,m,c)\). (def. 1.4)}
The same Mean Hypothesis score can also be expressed as an average across all 10 models (m) that comprise the council (def 1.5):
\EquationBox{Grand Mean Hypothesis Score}{\(\displaystyle \bar{S}_h(s,c)=\frac{1}{10}\sum_{m=1}^{10}\bar{S}_h(s,m,c)\)}{Mean hypothesis score across all models for a given \((s,c)\). (def. 1.5)}
We calculated pairwise mean-square differences for non-missing subset of the scores across hypotheses and contexts (all of the evaluations had at least 3 factors scored for each set, def 1.6):
\EquationBox{Mean-Square Difference (MSD)}{\(\displaystyle \operatorname{MSD}(A,B)=\frac{1}{6}\sum_{c\in\{\mathrm{LO},\mathrm{GO}\}}\sum_{h=1}^{3}\bigl(A(c,h)-B(c,h)\bigr)^2\)}{Squared difference between two hypothesis-score arrays. (def. 1.6)}
To compute the statistical significance of these scores and their distances, we performed random permutation tests, where scores were randomly permuted across the table either across all dimensions (full shuffle) or across a subset of dimensions (e.g., model, context, and factor fixed but randomizing study, def. 1.7, 1.8):
\EquationBox{Permutation Operator}{\(\pi:\Omega\to\Omega\) (def. 1.7)}{A random permutation sampled uniformly, \(S_{|\Omega|}\), representing a discrete shuffle of the index space \(\Omega\).}
\EquationBox{Full Shuffle}{\(\tilde{S}(i)=S(\pi(i)),\quad i=(s,m,c,f)\) (def. 1.8)}{A global shuffle where \(\pi\) permutes the entire flattened multi-dimensional index space simultaneously.}
For the hypothesis shuffle, we kept the model and the study fixed, but shuffled context and factors of hypotheses. (def. 1.9):
\EquationBox{Hypothesis Shuffle}{\(\tilde{S}(s,m,j)=S(s,m,\pi_{s,m}(j)),\quad j=(c,f)\) (def. 1.9)}{A particular permutation applied to the context-factor indices \((c,f)\) for each fixed study and model.}
\subsection{Agent Consistency (AC)}\label{agent-consistency-ac}

\emph{Agent Consistency (AC)} quantifies pairwise agreement between models, showing the relationship between the models in the council (equation 1.1). AC evaluates how closely individual models align in their textual interpretations across all studies (s):
\EquationBoxWideLeft{Agent Consistency (Pairwise)}{\(\displaystyle \operatorname{AC}(m_i,m_j)=\frac{1}{31}\sum_{s=1}^{31}\operatorname{MSD}\bigl(\bar{S}_h(s,m_i,\cdot),\bar{S}_h(s,m_j,\cdot)\bigr)\)}{Agreement between two models across studies. (eqn. 1.1)}
Averaging AC across all possible unique model combinations (for 10 models, 10*9/2=45) yields (equation 1.2), which is plotted as dotted lines in the Figures and indicated as AC:
\EquationBox{Grand Agent Consistency}{\(\displaystyle \operatorname{AC}=\frac{1}{45}\sum_{1\le i<j\le 10}\operatorname{AC}(m_i,m_j)\)}{Average consistency across all model pairs. (eqn. 1.2)}
\subsection{Literature Consistency (LC)}\label{literature-consistency-lc}

\emph{Literature Consistency} measures the study-study Mean-Squared Distance across the corpus, calculating how closely different publications align within the hierarchical predictive coding hypothesis space (equation 2.1):
\EquationBoxWideLeft{Literature Consistency (Pairwise)}{\(\displaystyle \operatorname{LC}(s_i,s_j)=\operatorname{MSD}\bigl(\bar{S}_h(s_i,\cdot),\bar{S}_h(s_j,\cdot)\bigr)\)}{Agreement between two studies. (eqn. 2.1)}
Averaging LC across all possible unique study combination (for 31 studies ((31*30)/2)=465) yields (equation 2.2), which is plotted as dotted lines in the Figures and indicated as LC:
\EquationBox{Grand Literature Consistency}{\(\displaystyle \operatorname{LC}=\frac{1}{465}\sum_{1\le i<j\le 31}\operatorname{LC}(s_i,s_j)\)}{Average consistency across study pairs. (eqn. 2.2)}
\subsection{Literature-Agent Consistency (LAC)}\label{literature-agent-consistency-lac}

The \emph{Literature-Agent Consistency (LAC)} represents a combined consistency score considering both models (m) and studies (s). Since AC,LC and LAC are all distance-based, lower values indicate greater consistency. Calculating LAC with respect to studies yields equation 3.1:
\EquationBox{LAC (Literature)}{\(\displaystyle \operatorname{LAC}_{\mathrm{lit}}(s,m)=\operatorname{MSD}\bigl(\bar{S}_h(s,m,\cdot),\bar{S}_h^{(-m)}(s,\cdot)\bigr)\)}{Model vs. other models for same study. (eqn. 3.1)}
Calculating LAC with respect to models yields equation 3.2:
\EquationBox{LAC (Agent)}{\(\displaystyle \operatorname{LAC}_{\mathrm{agent}}(s,m)=\operatorname{MSD}\bigl(\bar{S}_h(s,m,\cdot),\bar{S}_h^{(-s)}(m,\cdot)\bigr)\)}{Model vs. other studies for same model. (eqn. 3.2)}
Using equation 3.1 and 3.2 to calculate grand LAC, a distance-based scalar value used to represent overall agreement between literature and agents, in equation 3.3:
\EquationBox{Grand LAC}{\(\displaystyle \operatorname{LAC}_{\mathrm{grand}}=\frac{1}{2|S||M|}\sum_{s\in S}\sum_{m\in M}\left[\operatorname{LAC}_{\mathrm{lit}}(s,m)+\operatorname{LAC}_{\mathrm{agent}}(s,m)\right]\)}{Combined literature-agent consistency aggregated across all studies and models. (eqn. 3.3)}

\subsection{Statistical analysis}\label{statistical-analysis}
All summary statistics were computed from the validated model-study-context-factor score table. Scores marked as undefined were excluded from factor-level means rather than treated as zero. Hypothesis-level scores were computed by averaging non-missing factor scores within each hypothesis, then aggregating across models or studies as specified by the equations. Pairwise distances were quantified using mean-square difference over the local and global contexts and the three hypothesis dimensions. Local-versus-global context comparisons were evaluated as paired comparisons over matched study-model units or model-level aggregates, as indicated in the relevant analysis. The manuscript reports nominal alpha thresholds for the comparisons. The analysis code and exported score tables provide the basis for reproducing the reported comparisons, permutation references, and summary statistics.

\section*{Data availability}\label{data-availability}
The code, glossary, prompt templates, public-safe configuration placeholders, reproducibility notes, and citation metadata are available at https://github.com/HNXJ/mllm-public. Model outputs and derived score tables required to reproduce the analyses are available from the corresponding author on reasonable request and are planned for archival deposition with the public repository.

\section*{Code availability}\label{code-availability}
Analysis and pipeline code are available at https://github.com/HNXJ/mllm-public. The repository includes the HPC-36 predictive-coding glossary, prompt/instruction files, model/runtime documentation, public-safe configuration placeholders, reproducibility notes, and citation metadata.

\section*{Author contributions}\label{author-contributions}
H.N. designed the pipeline, implemented the analyses, curated the ontology, and drafted the manuscript. A.M., J.S.-S., and A.M.B. contributed to conceptual framing, interpretation, supervision, and manuscript revision. All authors reviewed the manuscript.

\section*{Competing interests}\label{competing-interests}
The authors declare no competing interests.

\section*{Ethics declarations}\label{ethics-declarations}
This study analyzed published literature and model-generated outputs only. No new human participant data, human tissue, live vertebrate animal experiments, or clinical intervention data were collected.

\section*{Additional information}\label{additional-information}
Large language models were used as scoring agents within the described local evaluation pipeline. They were not assigned authorship. The authors are responsible for the design, analysis, interpretation, and submitted text.

\clearpage
\appendix
\section*{Supplementary Information}
\addcontentsline{toc}{section}{Supplementary Information}

\section*{Supplementary Fig. S1}
\textbf{Supplementary Fig. S1:} The evaluation pipeline flows from left to right in two independent streams (A to B to C and 1 to 2 to 3). The A to B to C pathway establishes the instructions, canonical glossary, and prompt for the models. The 1 to 2 to 3 pathway ingests the paper and combines figures with text to provide a single text object for the models. Steps 5 and 6 loop between models and papers until all evaluations are completed. The final step 7 quantifies and visualizes the evaluations.

\begin{center}
\includegraphics[width=0.86\linewidth,keepaspectratio]{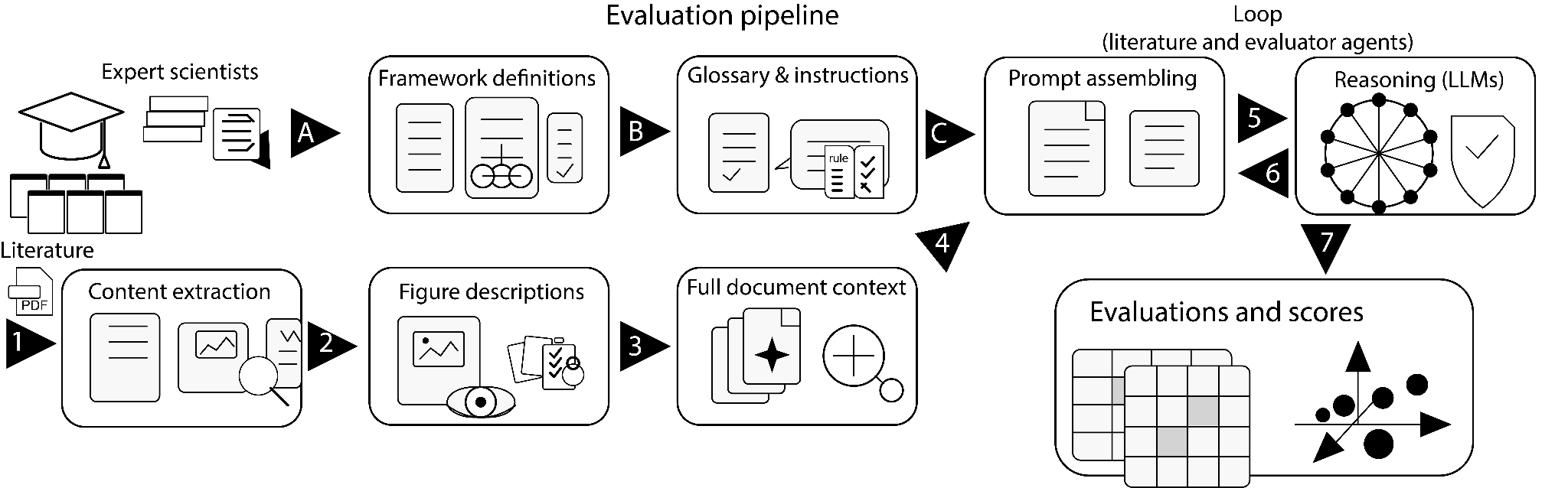}
\end{center}

\section*{Supplementary Tables}

\begin{longtable}[]{@{}
  >{\raggedright\arraybackslash}p{(\columnwidth - 6\tabcolsep) * \real{0.12}}
  >{\raggedright\arraybackslash}p{(\columnwidth - 6\tabcolsep) * \real{0.25}}
  >{\raggedright\arraybackslash}p{(\columnwidth - 6\tabcolsep) * \real{0.49}}
  >{\raggedright\arraybackslash}p{(\columnwidth - 6\tabcolsep) * \real{0.14}}
@{}}
\toprule\noalign{}
\begin{minipage}[b]{\linewidth}\raggedright
Factor Number
\end{minipage} & \begin{minipage}[b]{\linewidth}\raggedright
Factor Name
\end{minipage} & \begin{minipage}[b]{\linewidth}\raggedright
Definition
\end{minipage} & \begin{minipage}[b]{\linewidth}\raggedright
Context Scope
\end{minipage} \\
\midrule\noalign{}
\endhead
\bottomrule\noalign{}
\endlastfoot
1 & Subtractive Inhibition (SST) & Linear input suppression via somatic inhibition (Somatostatin-mediated). & LO+GO \\
2 & Divisive Inhibition (PV) & Multiplicative gain reduction (Parvalbumin-mediated). & LO+GO \\
3 & Inhibition (GABA) & Chloride-mediated inhibition for prediction suppression. & LO+GO \\
4 & Habituation to Sequence & Habituation suppresses local oddball response relative to a novel local oddball. & LO \\
5 & Synaptic Depression (Adaptation) & Passive fatigue of synaptic efficacy due to repetition of stimulus. & LO \\
6 & Activity Suppression & Reduction in firing rates for expected (predictable) stimuli. & LO+GO \\
7 & Selective Sharpening & Signal-to-noise increase for predictable stimuli, where noise is selectively suppressed to highlight relevant information. & LO+GO \\
8 & Alpha/Beta Mediated Suppression & Low-frequency bands (8-30Hz) associated with predictions, inhibiting prediction error signals. & LO+GO \\
9 & VIP-Mediated Disinhibition & VIP-mediated inhibition of SST/PV neurons, leading to disinhibited pyramidal activity during prediction error. & LO+GO \\
10 & Precision Weighting (Gain) & Top-down amplification of error units, leading to disinhibited gain in response to attended or highly precise stimuli. & LO+GO \\
11 & E/I Balance Shift & Dynamic adjustment toward higher inhibition for predictable stimuli, leading to suppressed neural firing. & LO+GO \\
12 & Omission Response & A generated internal signal due to absence of expected input & LO+GO \\
\end{longtable}

\textbf{Supplementary Table S1:} The factors of Hypothesis 1 (Predictive Suppression). LO = local oddball. GO = global oddball.

\begin{longtable}[]{@{}
  >{\raggedright\arraybackslash}p{(\columnwidth - 6\tabcolsep) * \real{0.12}}
  >{\raggedright\arraybackslash}p{(\columnwidth - 6\tabcolsep) * \real{0.25}}
  >{\raggedright\arraybackslash}p{(\columnwidth - 6\tabcolsep) * \real{0.49}}
  >{\raggedright\arraybackslash}p{(\columnwidth - 6\tabcolsep) * \real{0.14}}
@{}}
\toprule\noalign{}
\begin{minipage}[b]{\linewidth}\raggedright
Factor Number
\end{minipage} & \begin{minipage}[b]{\linewidth}\raggedright
Factor Name
\end{minipage} & \begin{minipage}[b]{\linewidth}\raggedright
Definition
\end{minipage} & \begin{minipage}[b]{\linewidth}\raggedright
Context Scope
\end{minipage} \\
\midrule\noalign{}
\endhead
\bottomrule\noalign{}
\endlastfoot
13 & Feedforward Deviance Detection & Detection of mismatch signals that explicitly propagate in the feedforward (ascending) direction to higher cortical areas. & LO+GO \\
14 & Feedforward AMPA & Fast excitatory drive mediated by AMPA receptors, specifically conveying prediction-error signals in the feedforward direction. & LO+GO \\
15 & Feedforward NMDA & Voltage-dependent amplification (bursting) of error signals, facilitating their robust feedforward propagation through the hierarchy. & LO+GO \\
16 & Feedforward Ascending Gamma & Rhythmic synchronization (30-90Hz) that temporally packages prediction-error signals for efficient feedforward transmission. & LO+GO \\
17 & Absence of Feedback Error & The functional requirement that the prediction-error signal be feedforward-directed, distinguishing it from local or feedback signals. & LO+GO \\
18 & Feedforward Non-local Supragranular Activity (L2/3) & Error-signaling neurons in Layers 2/3 projecting into the next area, demonstrating feedforward signal propagation. & LO+GO \\
19 & Feedforward Non-local Granular Activity (L4) & The arrival of prediction error signals in Layer 4, non-local representing the canonical entry point for feedforward propagation into the cortical column. & LO+GO \\
20 & Feedforward Non-local Directed Connectivity & Directional measures (Granger Causality/Transfer Entropy) confirming the directed feedforward propagation of prediction error from lower to higher areas. & LO+GO \\
21 & Feedforward Non-local Activation & Functional activation patterns specifically tracing the feedforward propagation path through the hierarchy. & LO+GO \\
22 & Ascending Latency Shift & Systematic increases in response latency that characterize the feedforward propagation of prediction error through successive levels of the hierarchy. & LO+GO \\
23 & Feedforward Error Propagation & Explicit experimental evidence that the prediction error signal is transmitted primarily via feedforward propagation pathways. & LO+GO \\
24 & Subcortical Feedforward Relaying & Relaying or generation of prediction error signals by subcortical structures (Thalamus/BG), initiating or sustaining their feedforward propagation to the cortex. & LO+GO \\
\end{longtable}

\textbf{Supplementary Table S2:} The factors of Hypothesis 2 (Feedforward Error Propagation)

\begin{longtable}[]{@{}
  >{\raggedright\arraybackslash}p{(\columnwidth - 6\tabcolsep) * \real{0.12}}
  >{\raggedright\arraybackslash}p{(\columnwidth - 6\tabcolsep) * \real{0.25}}
  >{\raggedright\arraybackslash}p{(\columnwidth - 6\tabcolsep) * \real{0.49}}
  >{\raggedright\arraybackslash}p{(\columnwidth - 6\tabcolsep) * \real{0.14}}
@{}}
\toprule\noalign{}
\begin{minipage}[b]{\linewidth}\raggedright
Factor Number
\end{minipage} & \begin{minipage}[b]{\linewidth}\raggedright
Factor Name
\end{minipage} & \begin{minipage}[b]{\linewidth}\raggedright
Definition
\end{minipage} & \begin{minipage}[b]{\linewidth}\raggedright
Context Scope
\end{minipage} \\
\midrule\noalign{}
\endhead
\bottomrule\noalign{}
\endlastfoot
25 & Canonical Microcircuit Ubiquity & The ubiquitous presence of the L2/3 Error and L5/6 Prediction motif repeating across most or all levels of the cortical hierarchy. & LO+GO \\
26 & Hierarchical Mechanism Invariance & The ubiquitous nature of the predictive coding mechanism, functioning similarly in the hierarchy from V1 to PFC. & LO+GO \\
27 & Hierarchical Activity Ubiquity & The ubiquitous presence of prediction-error activity, detectable across all or most levels of the system's brain hierarchy. & LO+GO \\
28 & Hierarchical CSD Ubiquity & The ubiquity of current source density profiles, which show consistent laminar patterns across most or all hierarchical levels. & LO+GO \\
29 & Cross-Scale Hierarchical Ubiquity & The ubiquity of effects observable across scales (single units to LFP) throughout most levels of the cortical hierarchy. & LO+GO \\
30 & Hierarchical Presence (V1-PFC) & The ubiquity of the mechanism across all levels of the hierarchy, such as low-level (V1), mid-level (V4), and high-level (PFC) areas. & LO+GO \\
31 & Cross-Modal Ubiquity & Presence of the mechanism across multiple sensory modalities (visual, auditory, somatosensory) throughout the hierarchy. & LO+GO \\
32 & Interspecies Hierarchical Ubiquity & Conservation of the hierarchical predictive coding mechanism across different species (e.g., mouse vs.~primate). & LO+GO \\
33 & Temporal Stability of Ubiquity & The consistent and non-transient presence of these mechanisms across the hierarchy over long recording sessions. & LO+GO \\
34 & Hierarchical Order Stability & Evidence that the predictive mechanism is ubiquitous and not localized to a single hierarchical pole. & LO+GO \\
35 & Population-Wide Ubiquity & Consistency of the mechanism across diverse neural populations and cell types throughout the hierarchy. & LO+GO \\
36 & State-Independent Ubiquity & Presence of the mechanism across different brain states (e.g., wakefulness, sleep, anesthesia) throughout the hierarchy. & LO+GO \\
\end{longtable}

\textbf{Supplementary Table S3:} The factors of Hypothesis 3 (Ubiquity)

\begin{longtable}[]{@{}
  >{\raggedright\arraybackslash}p{(\columnwidth - 4\tabcolsep) * \real{0.20}}
  >{\raggedright\arraybackslash}p{(\columnwidth - 4\tabcolsep) * \real{0.43}}
  >{\raggedright\arraybackslash}p{(\columnwidth - 4\tabcolsep) * \real{0.37}}
@{}}
\toprule\noalign{}
\begin{minipage}[b]{\linewidth}\raggedright
Term/Context
\end{minipage} & \begin{minipage}[b]{\linewidth}\raggedright
Description
\end{minipage} & \begin{minipage}[b]{\linewidth}\raggedright
Definition (notation)
\end{minipage} \\
\midrule\noalign{}
\endhead
\bottomrule\noalign{}
\endlastfoot
Oddball & A surprising event, often a mismatch & S = \{B\} given E{[}S{]} = \{A\} \\
Local Oddball (LO) & An oddball that is immediate or short-term, often confounded by low-level adaptation mechanisms. & S = \{AB\} given E{[}S{]} = \{AA\} \\
Global Oddball (GO) & An oddball that is not LO, often contextual, that is not caused by adaptation mechanisms. & S = \{AA\} given E{[}S{]} = \{AB\} \\
\end{longtable}

\textbf{Supplementary Table S4:} Contexts ; S=\{.\} represents the sensory input, E{[}S{]} represents the expected sensory input. When S=\{A\} and E{[}S{]} = \{A\} ; the prediction error is zero (match). When S=\{B\} and E{[}S{]} = \{A\} ; the prediction error is non-zero (mismatch). LO and GO differ by the temporal scale of expectation.

\begin{longtable}[]{@{}
  >{\raggedright\arraybackslash}p{(\columnwidth - 4\tabcolsep) * \real{0.06}}
  >{\raggedright\arraybackslash}p{(\columnwidth - 4\tabcolsep) * \real{0.23}}
  >{\raggedright\arraybackslash}p{(\columnwidth - 4\tabcolsep) * \real{0.71}}
@{}}
\toprule\noalign{}
\begin{minipage}[b]{\linewidth}\raggedright
\#
\end{minipage} & \begin{minipage}[b]{\linewidth}\raggedright
Study
\end{minipage} & \begin{minipage}[b]{\linewidth}\raggedright
Short Description
\end{minipage} \\
\midrule\noalign{}
\endhead
\bottomrule\noalign{}
\endlastfoot
1 & Attinger et al., 2017\cite{ref60} & Integrating non-visual (motor) inputs into V1 to generate mismatch signals during locomotion. \\
2 & Bakhtiari et al., 2021\cite{ref61} & Alignment between deep neural networks and biological vision across different predictive goals. \\
3 & Bastos et al., 2012\cite{ref14} & Canonical microcircuits for predictive coding; proposing distinct roles for superficial and deep layers. \\
4 & Bastos et al., 2020\cite{ref24} & Laminar and directional flow of prediction/error signals in primate visual and prefrontal areas. \\
5 & Bekinschtein et al., 2009\cite{ref62} & Neural markers of conscious perception using hierarchical auditory oddball tasks (fMRI/EEG). \\
6 & Chao et al., 2018\cite{ref26} & Global distribution of prediction errors across the primate brain using large-scale ECoG mapping. \\
7 & Friston et al., 2010\cite{ref10} & The Free Energy Principle: a unified framework for brain function and biological self-organization. \\
8 & Furutachi et al., 2024\cite{ref63} & Cooperative thalamocortical circuit mechanism for sensory prediction errors. \\
9 & Garrett et al., 2020\cite{ref64} & Emergence of prediction errors in the mouse visual cortex for surprises in sequence patterns. \\
10 & Greedy et al., 2022\cite{ref65} & Uses a new model, Bursting Cortico-Cortical Networks to approximate error backpropagation in a biological network of neurons. \\
11 & Hertag et al., 2020\cite{ref66} & Role of inhibitory interneuron types (PV and SST) in shaping cortical prediction errors. \\
12 & Jiang et al., 2024\cite{ref67} & Integrating hierarchical predictive coding for sequence learning. \\
13 & Keller et al., 2012\cite{ref68} & Sensorimotor prediction errors in V1; neurons responding to mismatches between movement and visual flow. \\
14 & Keller et al., 2018\cite{ref23} & Review of evidence of predictive processing and its implementation in mammalian cortical circuits. \\
15 & Kiebel et al., 2008\cite{ref69} & Hierarchical temporal dynamics; suggesting slower dynamics represent higher-level predictions. \\
16 & Lao et al., 2023\cite{ref70} & Provides evidence that top-down feedback is necessary for generating prediction error signals from omitted-tones. \\
17 & Lee et al., 2025\cite{ref71} & Computational modeling of PV/SST/VIP inhibitory interneurons and their dynamics during predictive processing \\
18 & Mikulasch et al., 2023\cite{ref72} & How predictive coding might utilize dendritic error computations \\
19 & Nejad et al., 2025\cite{ref73} & A self-supervised computational learning theory that aligns with the layer-specific neuronal observations in primary visual cortex \\
20 & Payeur et al., 2021\cite{ref74} & Burst-based firing enables the multiplexing of feedforward and feedback signals for predictive coding. \\
21 & Rao et al., 2024\cite{ref75} & Generalized predictive coding and sensorimotor theory of neocortex \\
22 & Rao\&Ballard, 1999\cite{ref59} & Seminal model of hierarchical predictive coding explaining extra-classical receptive field effects in V1. \\
23 & Sacramento et al., 2018\cite{ref76} & Dendritic model explaining how neurons learn top-down predictions via local error signals. \\
24 & Spratling et al., 2008\cite{ref77} & PC-BC model: implementing predictive coding within a biased competition framework for object recognition. \\
25 & Spratling et al., 2010\cite{ref78} & Predictive coding as a model of response properties in cortical area V1 \\
26 & Srinivasan et al., 1982\cite{ref7} & Predictive coding in the retina; earliest formalization of removing redundancy from natural images. \\
27 & VanDerveer et al., 2021\cite{ref79} & Somatostatin-Positive interneurons in neuro-oscillatory deficits \\
28 & Wacongne et al., 2012\cite{ref80} & Spiking neural network reproducing the auditory mismatch negativity (MMN) through predictive mechanisms. \\
29 & Wacongne et al., 2011\cite{ref52} & MEG study showing the human brain detects violations of abstract hierarchical rules in sound sequences. \\
30 & Westerberg et al., 2025\cite{ref34} & A recent study using Multi-Area, High-Density, Laminar Neurophysiology (MaDeLaNe) recordings in mice and monkeys to investigate global-local oddballs \\
31 & Yamins et al., 2014\cite{ref81} & Using goal-driven deep hierarchies (HCNNs) to predict neural responses across the visual cortex. \\
\end{longtable}

\textbf{Supplementary Table S5:} A total of 31 papers were considered in the current work. 25 papers were chosen as representative of the predictive coding literature, based on Westerberg et al.\cite{ref34} and span contributions ranging across theoretical, empirical, and computational sub-fields of predictive coding (PC). Additional papers (Bakhtiari et al., 2021; Greedy et al., 2022; Payeur et al., 2021; Sacramento et al., 2018; Yamins et al., 2014 ) were included as they represent an adjacent field of computational neuroscience where alternative mechanisms for prediction error propagation have been proposed. An additional paper, Garrett et al., 2020, was also included to consider predictive coding mechanisms related to omissions, as these responses may be partially overlapping or distinct from classical PC mechanisms.

\begin{longtable}[]{@{}
  >{\raggedright\arraybackslash}p{(\columnwidth - 4\tabcolsep) * \real{0.38}}
  >{\raggedright\arraybackslash}p{(\columnwidth - 4\tabcolsep) * \real{0.10}}
  >{\raggedright\arraybackslash}p{(\columnwidth - 4\tabcolsep) * \real{0.52}}
@{}}
\toprule\noalign{}
\begin{minipage}[b]{\linewidth}\raggedright
Model Name (handle-mlx)
\end{minipage} & \begin{minipage}[b]{\linewidth}\raggedright
Size
\end{minipage} & \begin{minipage}[b]{\linewidth}\raggedright
Release Date and info
\end{minipage} \\
\midrule\noalign{}
\endhead
\bottomrule\noalign{}
\endlastfoot
gemma-3-27b-it & 27B & Google DeepMind, 2025-03 \\
gemma-4-31b-it & 31B & Google DeepMind, 2026-04 \\
deepseek-r1-distill-llama-70b & 70B & DeepSeek-AI / Meta AI (Llama Base), 2025-01 \\
gpt-oss-claude-4.5-sonnet & 20B & Community Distillation (Teacher: Claude 4.5), 2026-03 \\
mistral-nemo-12b-thinking & 12B & Mistral AI \& NVIDIA, 2024-07 \\
olmo-3-32b-think & 32B & Allen AI (Ai2), 2025-12 \\
phi-4-reasoning-plus & 14B & Microsoft, 2025-03 \\
qwen3-14b-gemini-3-pro & 14B & Qwen (Community Fine-tune), 2025-12 \\
qwen3-14b-claude-4.5-sonnet & 14B & Qwen (Community Fine-tune), 2025-12 \\
qwen3.5-40b-claude-4.5-opus & 40B & Qwen (Community Fine-tune), 2026-01 \\
\end{longtable}

\textbf{Supplementary Table S6:} Open-weight LLMs used in this work. All models were running with MLX-LM (the python package for running LLMs locally on Apple's silicon hardware, M-series).

\end{document}